\documentclass[openacc]{rstransa}%%%%where rstrans is the template name
\usepackage{bm}
\usepackage{xcolor}
\usepackage{pdflscape}
\usepackage{mciteplus}
\titlehead{Research}

\begin{document}

%%%% Article title to be placed here
\title{The many faces of rotating quantum turbulence}

\author{%%%% Author details
Julian Amette Estrada$^{1,2}$, Marc E. Brachet$^{3}$ and Pablo D. Mininni$^{1,2}$}

%%%%%%%%% Insert author address here
\address{$^{1}$Universidad de Buenos Aires, Facultad de Ciencias Exactas y Naturales, Departamento de Física, Ciudad Universitaria, 1428 Buenos Aires, Argentina\\
$^{2}$CONICET - Universidad de Buenos Aires, Instituto de F\'{\i}sica Interdisciplinaria y Aplicada (INFINA), Ciudad Universitaria, 1428 Buenos Aires, Argentina,\\
$^{3}$Laboratoire de Physique de l'{E}cole {N}ormale {S}up\'erieure, ENS, Universit\'e PSL, CNRS, Sorbonne Universit\'e, Universit\'e de Paris, F-75005 Paris, France.}

%%%% Subject entries to be placed here %%%%
\subject{xxxxx, xxxxx, xxxx}

%%%% Keyword entries to be placed here %%%%

%%%% Insert corresponding author and its email address}
\corres{Julian Amette Estrada\\
\email{julianamette@df.uba.ar}}

%%%% Abstract text to be placed here %%%%%%%%%%%%
\begin{abstract}
Quantum turbulence shares many similarities with classical turbulence in the isotropic and homogeneous case, despite the inviscid and quantized nature of its vortices. However, when quantum fluids are subjected to rotation, their turbulent dynamics depart significantly from the classical expectations. We explore the phenomenology of rotating quantum turbulence, emphasizing how rotation introduces new regimes with no classical analogs. We review recent theoretical, experimental, and numerical developments, and present new numerical results that map out distinct dynamical regimes arising from the interplay of rotation, quantization, non-linearities, and condensed matter regimes. In particular, we show the importance of distinguishing the dynamics of rotating quantum fluids in the slowly rotating, rapidly rotating, and low Landau level regimes. The findings have implications for the dynamics of liquid helium, atomic Bose-Einstein condensates, and neutron stars, and show how rotating quantum fluids can serve as a unique platform bridging turbulence theory and condensed matter physics revealing novel states of out-of-equilibrium quantum matter.
\end{abstract}
%%%%%%%%%%%%%%%%%%%%%%%%%%%

%%%%%%%%%% Insert the texts which can accomdate on firstpage in the tag "fmtext" %%%%%

%\begin{fmtext}
%\end{fmtext}

\maketitle
%%%%%%%%%%%%%%% End of first page %%%%%%%%%%%%%%%%%%%%%

\section{Introduction}

Turbulence in quantum fluids---such as the chaotic and disordered dynamics observed in superfluid helium and atomic Bose–Einstein condensates (BECs)---has evolved from a curious theoretical prediction \cite{Feynman1955} into an active topic at the intersection of fluid dynamics and quantum physics. While many statistical features of isotropic and homogeneous quantum turbulence bear resemblance to their classical counterparts, important distinctions arise from the quantization of circulation and the inviscid nature of the superfluid phase in the quantum case \cite{Skrbek2012, Henn2009, Nore1997}. This impacts, for example, in the statistical properties of the velocity field and the occurrence of extreme values of the velocity in the vicinity of quantum vortices. But in spite of this, and while in some cases it may lead to different scaling properties, in many cases the scaling laws followed by classical and quantum turbulence, and the general properties of turbulence, are the same \cite{Barenghi2023}. However, these differences become more pronounced, giving rise to new regimes, when external constraints break the flow isotropy, as is the case in the presence of rotation.

Rotating turbulence in classical fluids has been extensively studied in recent years, motivated by its relevance in industrial and geophysical contexts \cite{ClarkDiLeoni2020, vanKan2021, Alexakis_2024}. It is known to exhibit a rich phenomenology, which includes the development of anisotropy, the appearance of large-scale columnar vortex structures, and the suppression of small-scale energy transfer \cite{Cambon_1997, BELLET2006, Sen_2012, Scott2014}. Yet, when rotation is introduced in quantum fluids, new and fundamentally different regimes emerge---regimes that do not have classical analogues \cite{AmetteEstrada2022a}. In these systems, the combination of rotation and quantization enforces the formation of a dense lattice of quantum vortices, which introduces long-range order, and fundamentally alters the dynamics of turbulence \cite{Coddington2003, AmetteEstrada2022a, Peretti2023, Mkinen2023}. These effects lead to unique states of out-of-equilibrium quantum matter, where familiar concepts from classical turbulence theory and condensed matter physics must be simultaneously considered to explain the flow behaviour. This opens the possibility of connecting rotating quantum turbulence to others systems such as superconductors \cite{Severino_2022}, and, following recent work, to dark matter and neutron stars where solutions of the Gross-Pitaevskii equation can provide novel insights \cite{Shukla2024, Liu2025, Sivakumar2025}. 

This paper aims to bridge these two domains by surveying recent progress on rotating quantum turbulence, while also presenting new simulations that explore the rich variety of dynamical regimes accessible in this system. In Sec.~\ref{sec:review} we review previous theoretical, laboratory, and numerical studies of rotating quantum fluids. While we discuss some recent experiments using liquid helium, we mostly focus on gaseous BECs, and thus we introduce the theoretical framework of how to describe these systems using the Gross-Pitaevskii model. We then put a special emphasis on the available measurements or numerical simulations of the out-of-equilibrium regime of rotating condensates. The reader interested in a general review of the equilibrium properties of rotating condensates is pointed to \cite{Fetter_2009}. The available results reported so far in rotating quantum turbulence show a myriad of different regimes. Then, in Sec.~\ref{sec:results}, we present new numerical simulations that explore systematically the effect of varying the rotation rate compared with the trapping potential as well as other parameters in the system. We show that different parameter regimes---including rotation rate, interaction strength, and system size---give rise to qualitatively distinct flow behaviours that challenge classical intuitions about turbulence. In particular, we show that turbulence is strongly affected by whether the system is in the slowly rotating, rapidly rotating, or low Landau level regime. These findings underline the importance of incorporating tools and concepts from both turbulence theory and quantum many-body physics to capture the full phenomenology of rotating quantum flows.

\section{Physics of rotating Bose-Einstein condensates \label{sec:review}}

\subsection{A brief history of condensation}

Since the introduction in 1924 of Bose-Einstein statistics, developed by Satyendra Nath Bose and Albert Einstein \cite{bose1924plancks, Einstein1924, Einstein1925}, the phenomenon of condensation presented itself as a strange and abstract idea. What we now see as a macroscopic manifestation of a quantum phenomenon was much more difficult to make sense at a time when quantum theory was yet not fully developed. The scepticism around the idea started to fade later when Fritz London \cite{london1938lambda} and László Tisza \cite{tisza1938transport} suggested a connection between Bose-Einstein condensation and the $\lambda$ transition observed in $^4$He, despite liquid helium being a strongly interacting system. As time passed, the idea of observing quantum behaviour of matter at macroscopic scales was accepted, with superconductivity and superfluidity providing some of the its best known examples.

In spite of the early success of the theory, the cases of superconductors or of superfluid $^4$He lie beyond its range of validity. As an example, strong interactions in helium leads to a reduced amount of particles in the zero momentum state, even at zero temperature \cite{Pethick2008}. This motivated the pursue of obtaining a BEC in a dilute atomic gas, which implied a great experimental effort that lasted for decades. The first gaseous BEC was finally achieved in the laboratory in 1995 \cite{Anderson1995}.

Since then, several experiments with quantum gases or liquid helium have been carried out. Shortly after the first atomic BEC was observed, quantum vortices were generated and studied in this system, which had also been studied before for several decades in superfluid helium \cite{VINEN1958, Yarmchuk1979}. And in 2009, observations of quantum turbulence were reported in oscillating atomic BECs \cite{Henn2009}, after several decades of studies of this phenomenon in superfluid helium (see, e.g., \cite{Skrbek2012}). This sparked a renewed interest in the dynamics of out-of-equilibrium of BECs and superfluids, and in the evolution of disordered and chaotic flows in these systems \cite{Nore1997, Barenghi2014, Barenghi2023}.

The study of quantum turbulence remains an important field with several recent theoretical and experimental advances, and a focus on understanding its particular physics and its relation with its classical counterpart \cite{Skrbek2012, Barenghi2023}. In recent years, fuelled by advances in experiments and in available technologies, as well as advances in numerical simulations, research has also considered the case of rotating BECs and superfluids. New experiments focused either in near equilibrium states (including the development of vortex lattices and collective oscillations), or in far from equilibrium turbulent regimes. Focusing on the description of rotating quantum gases, we next introduce the theoretical methods used to describe them, and the different condensed matter regimes and waves that can be excited in the system.

\subsection{The Gross-Pitaevskii equation}
\label{sec:GPE}

For dilute quantum gases, the dynamics of a large number of weakly interacting bosons in a BEC at zero temperature in the non-rotating case is given by the Gross-Pitaevskii equation (GPE)\cite{Pethick2008}:
\begin{equation}
    i \hbar \frac{\partial \psi}{\partial t}(\mathbf{r}',t) = \left[ -\frac{\hbar^2}{2m} \nabla'^2 + g \left| \psi (\mathbf{r}',t) \right|^2 + V(\mathbf{r}')  \right] \psi(\mathbf{r}',t),
    \label{eq: GPE}
\end{equation}
where $\psi$ is the order parameter of the condensate, $m$ is the boson mass, $g= 4 \pi a \hbar^2/m$, $a$ is the s-wave scattering length, $\textbf{r}'$ is the position in the laboratory reference frame, $\mathbf{p}' = -i \hbar \boldsymbol{\nabla}'$ is the momentum operator in this frame with $p'^2 = -\hbar^2 \nabla'^2$, and $V(\mathbf{r}',t)$ is an external potential. 

We are interested in the expression of GPE for a bosonic gas under a uniform rotation with angular velocity $\boldsymbol{\Omega}$, in the rotating reference frame. Positions in the rotating coordinate system are given by ${\bf r} =  R(\boldsymbol{\Omega},t) \mathbf{r}'$, where $R(\boldsymbol{\Omega},t)$ is the three-dimensional rotation matrix at time $t$. Given an order parameter with coordinates co-rotating with the gas, $\psi({\bf r},t)$, we must rotate it to obtain its equivalent in the laboratory, i.e., $\psi({\bf r}',t) = \psi(R^T(\boldsymbol{\Omega},t) {\bf r},t) = \hat{R}(\boldsymbol{\Omega},t) \psi({\bf r},t)$, where the super-index $T$ denotes a transposition, $\hat{R}(\boldsymbol{\Omega},t) = e^{-i \mathbf{J} \cdot \boldsymbol{\Omega} t/\hbar}$ is the rotation operator, and $\mathbf{J} = \mathbf{r} \times \mathbf{p}$ is the angular momentum operator. Equivalently, $\psi({\bf r},t) = \psi(R(\boldsymbol{\Omega},t) {\bf r}',t) = \hat{R}^\dagger(\boldsymbol{\Omega},t) \psi({\bf r}',t)$, or in other words, to obtain the order parameter in the rotating frame from the one in the laboratory frame we must ``unwind'' its rotation. From now on, in instances where its use does not lead to ambiguity or confusion, we will use  the shorthand notation $\psi' = \psi (\mathbf{r}',t)$ for the order parameter in the non-rotating reference frame, $\psi = \psi (\mathbf{r},t)$ in the rotating coordinates, and in general, for any function in these reference frames, $f' = f(\mathbf{r}',t)$ and $f = f(\mathbf{r},t)$.

The order parameter $\psi'$ satisfies the usual Gross-Pitaevskii equation. Applying $\hat{R}^{\dagger}(\boldsymbol{\Omega},t)$ to the left of Eq.~(\ref{eq: GPE}), and using $[\hat{R}^\dagger(\boldsymbol{\Omega},t) , \partial_t] = -(i/\hbar) \boldsymbol{\Omega} \cdot \mathbf{J} \hat{R}^{\dagger}(\boldsymbol{\Omega},t)$, $[\hat{R}^{\dagger}(\boldsymbol{\Omega},t),\nabla'^2]=0$, $\nabla'^2 = \nabla^2$, $|\psi'|^2 = |\psi|^2$, and $V' = \hat{R}(\boldsymbol{\Omega},t) V \hat{R}^\dagger(\boldsymbol{\Omega},t)$, we finally obtain the rotating Gross-Pitaevskii equation (RGPE) \cite{Fetter2008,AmetteEstrada2022a,AmetteEstrada2022b}:
\begin{equation}
    i \hbar \frac{\partial \psi}{\partial t}(\mathbf{r},t) = \left[ -\frac{\hbar^2}{2m} \nabla^2 + g \left| \psi (\mathbf{r},t) \right|^2 + V(\mathbf{r}) - \boldsymbol{\Omega} \cdot \mathbf{J} \right] \psi(\mathbf{r},t) ,
    \label{eq: RGPE}
\end{equation}
which describes a rotating condensate in the reference frame co-rotating with the gas. A rotating BEC can be studied using Eq.~(\ref{eq: GPE}) in the laboratory frame, with an initial configuration with net rotation or using a potential that generates a rotation of the gas, or in the rotating frame using Eq.~(\ref{eq: RGPE}). From a theoretical perspective, RGPE gives a better description of the system as the potentials confining or doing work on the system in the laboratory frame can be non-stationary.

We can provide a hydrodynamic description of the gas in both reference frames. In the non-rotating frame, the Madelung transformation
\begin{equation}
    \psi(\mathbf{r}',t)= \sqrt{\frac{\rho (\mathbf{r}',t)}{m}} \,\,e^{i S(\mathbf{r}',t)},
\end{equation}
where $\rho (\mathbf{r}',t)$ is the gas mass density and $S(\mathbf{r}',t)$ is the phase of the order parameter, allows us to write the gas dynamics using fluid-like quantities. A continuity equation for the mass density can be derived by multiplying Eq.~(\ref{eq: GPE}) by $\psi'^*$, where the star denotes the complex conjugate, and by subtracting the complex conjugate of the resulting equation, to obtain
\begin{equation}
    \frac{\partial \rho'}{\partial t} + \boldsymbol{\nabla}' \cdot \mathbf{j}' = 0 .
    \label{eq:cont}
\end{equation}
This equation allows us to define the momentum density,
\begin{equation}
    \mathbf{j}(\mathbf{r}',t) = - \frac{i \hbar}{2} \left( {\psi'}^* \boldsymbol{\nabla}' \psi' - \psi' \boldsymbol{\nabla}' {\psi'}^*\right), 
\end{equation}
such that the gas velocity is $\mathbf{v}' = \mathbf{j}'/\rho'$. This velocity can be written as
\begin{equation}
    \mathbf{v}' = \frac{\hbar}{m} \boldsymbol{\nabla}'S',
\end{equation}
which is well defined and irrotational as long as there are no topological defects in the gas---i.e., if $\psi'\neq 0$ everywhere. In the topological defects, the circulation of the velocity is quantized. This follows, as argued by Feynman \cite{Feynman1955}, from the fact that the order parameter must be single-valued. As a result, when going around a topological defect in a closed path, the total accumulated phase change can only be $\oint \boldsymbol{\nabla}' S' \cdot d\boldsymbol{\ell}' = 2 \pi n$,
with $n\in \mathbb{Z}$. This implies that
\begin{equation}
\Gamma = \oint \mathbf{v}' \cdot d\boldsymbol{\ell}' = \Gamma_0 n,
\end{equation}
where $\Gamma$ is the total circulation, and $\Gamma_0 = h / m$ is the quantum of circulation. When a single defect (i.e., a quantum vortex) is present, the fluid vorticity $\bm{\omega}' = \boldsymbol{\nabla}' \times \mathbf{v}'$ can be written as
\begin{equation}
    \boldsymbol{\omega}' = n \Gamma_0 \int ds \frac{d \mathbf{r}'_0}{ds} \delta^{(3)}(\mathbf{r}'-\mathbf{r}'_0(s)) ,
\end{equation}
where $\mathbf{r}'_0(s)$ is the coordinate of the $\psi'=0$ centreline, and $s$ is an arc-length parameter.

In the rotating reference frame, the Madelung transformation $\psi = \sqrt{\rho/m} \exp (i S)$ and Eq.~(\ref{eq: RGPE}) yield again a continuity equation as in Eq.~(\ref{eq:cont}), but with a momentum density
\begin{equation}
    \mathbf{j}(\mathbf{r},t) = - \frac{i \hbar}{2} \left( {\psi}^* \boldsymbol{\nabla} \psi - \psi \boldsymbol{\nabla} {\psi}^*\right) - \rho \bm{\Omega} \times \mathbf{r}.
    \label{eq: momentum density rot}
\end{equation}
This gives a gas velocity in the rotating frame
\begin{equation}
    \mathbf{v}_\textrm{rot} = \frac{\mathbf{j}}{\rho} = \frac{\hbar}{m} \boldsymbol{\nabla} S - \bm{\Omega} \times \mathbf{r} = \mathbf{v} - \bm{\Omega} \times \mathbf{r} ,
   \label{eq:vrot} 
\end{equation}
where $\mathbf{v} = (\hbar/m)\boldsymbol{\nabla} S$ is the same total velocity as $\mathbf{v}'$ (including the velocity of the rigid body rotation of the condensate) but just passively rotated. Thus, computing the gas velocity from the gradient of the phase of $\psi$ or $\psi'$ gives the total velocity in all cases, and both $\rho$ and $\rho'$, and $\mathbf{v}$ and $\mathbf{v}'$, are the same quantities under a passive rotation. Moreover, the fact that $\mathbf{v}$ and $\mathbf{v}'$ are irrotational everywhere except at singular points also has an important implication: The only way for the BEC to mimic a solid body rotation is by generating an array of quantum vortices, ordered in such a way that their coarse-grained velocity generates a net rotation.

Finally, in the absence of quantum vortices, both GPE and RGPE can be rewritten as equations akin to those of an isentropic, compressible, and irrotational fluid, with quantum and interaction corrections. This was done first for the linear Schrödinger equation by Madelung in 1927 \cite{Madelung1927}. For GPE in the laboratory reference frame, the equivalent procedure is done by applying the Madelung transformation to Eq.~(\ref{eq: GPE}). The imaginary part  of Eq.~(\ref{eq: GPE}) gives again Eq.~(\ref{eq:cont}), i.e., the continuity equation, while the gradient of the real part gives a modified Euler equation \cite{Nore1997}.  In the rotating reference frame, noting that $\mathbf{v}$ is irrotational but $\mathbf{v}_\textrm{rot}$ is not, this procedure gives
\begin{align}
    \frac{\partial \rho}{\partial t} + \boldsymbol{\nabla} \cdot (\rho \mathbf{v}_\textrm{rot}) & = 0 , \\
    \frac{\partial \mathbf{v}_\textrm{rot}}{\partial t} + \mathbf{v}_\textrm{rot} \cdot \boldsymbol{\nabla} \mathbf{v}_\textrm{rot} & = - \boldsymbol{\nabla} \left( \frac{g \rho}{m^2} -\frac{\hbar^2}{2m^2} \frac{\nabla^2 \sqrt{\rho}}{\sqrt{\rho}} \right) - 2 \boldsymbol{\Omega} \times \mathbf{v}_\textrm{rot} - \boldsymbol{\Omega} \times \boldsymbol{\Omega} \times \mathbf{r} + \mathbf{f} ,
\end{align}
where $g \rho/m^2$ is the pressure arising from the interactions between particles, $-\hbar^2 \nabla^2 \sqrt{\rho}/( 2 m^2 \sqrt{\rho})$ is the quantum pressure, $-2 \boldsymbol{\Omega} \times \mathbf{v}_\textrm{rot}$ is the Coriolis acceleration, $-\boldsymbol{\Omega} \times \boldsymbol{\Omega} \times\mathbf{r}$ is the centrifugal acceleration, and $\mathbf{f} = -\boldsymbol{\nabla} V/m$ is an external force per unit mass.

This formalism can be extended to consider BECs at finite temperature; for a discussion in rotating and non-rotating flows we refer the reader to  \cite{Proukakis2008, Berloff2014, AmetteEstrada2022b, Amette_2025}. The equations can also be extended to capture effects relevant in liquid helium, such as roton excitations \cite{Muller_2020}. Finally, RGPE shares similarities with the equations that describe type II superconductors under external magnetic fields. Taking the curl of Eq.~\eqref{eq: momentum density rot} gives $\boldsymbol{\nabla} \times \mathbf{j} = - 2 \rho \bm{\Omega}$, which is the second London equation \cite{london1935} under the changes $2 \rho \rightarrow n_s e^2/2$, where $n_s$ is the density of superconducting charges and $e$ the electron charge, and $\bm{\Omega} \rightarrow \bm{B}$, where $\bm{B}$ is the external magnetic field.

\subsection{Hamiltonian of the system}
\label{sec: hamiltonian}

We can write a Hamiltonian for a rotating gas of weakly interacting bosons. The formulation is useful to identify the different contributions to the system energy, and to see how rotation changes the equilibrium. We start again in the laboratory reference frame, with the Hamiltonian of GPE,
\begin{equation}
    \mathcal{H}_0' [\psi',\psi'^*] = \int d^3 r' \left[ \frac{\hbar^2}{2m} |\boldsymbol{\nabla}' \psi'|^2 + \frac{g}{2} |\psi'|^4 + V(\mathbf{r}') |\psi'|^2\right].
\end{equation}
In the rotating reference frame we must replace $\mathbf{r}'$ by $\mathbf{r}$ (i.e., $\psi'$ by $\psi$), and consider a new contribution to the Hamiltonian \cite{Fetter2008, AmetteEstrada2022b} to capture the last term on the r.h.s.~of RGPE in Eq.~(\ref{eq: RGPE}),
\begin{equation}
    \mathcal{H}_\textrm{rot}[\psi,\psi^*] = -  \int d^3r \, \psi^* \bm{\Omega} \cdot \mathbf{J} \psi. 
\end{equation}
The total Hamiltonian for the condensate in the rotating frame them is $\mathcal{H} = \mathcal{H}_0[\psi,\psi^*] + \mathcal{H}_\textrm{rot}[\psi,\psi^*]$. From the discussion in Sec.~2\ref{sec:GPE}, passive rotations do not affect volume averaged quantities. Taking this into account, and using the Madelung transformation, the total Hamiltonian for RGPE can be written as
\begin{equation}
    \mathcal{H}[\psi,\psi]= \int d^3 r' \left[ \frac{\rho}{2} |\mathbf{v}|^2 + \frac{\hbar^2}{2 m^2} \left| \boldsymbol{\nabla} \sqrt{\rho} \right|^2 + \frac{g \rho^2}{2m^2} + \frac{V(\mathbf{r})}{m} \rho - \psi^*  \bm{\Omega} \cdot \mathbf{J} \psi \right].
    \label{eq:HRGPE}
\end{equation}

The Hamiltonian in Eq.~(\ref{eq:HRGPE}) defines the energy in the rotating frame, formally only differing in the last term from the Hamiltonian in the laboratory frame. At this point we can decompose this Hamiltonian into several energy components of the form
\begin{eqnarray}
    E_k  &=& \left< \frac{\rho}{2} |{\bf v}|^2\right>,\\
    E_q  &=& \left< \frac{\hbar^2}{2m^2} \left| \boldsymbol{\nabla} \sqrt{\rho} \right|^2\right>,\\
    E_p  &=& \left< \frac{g \rho^2}{2m^2} \right>,\\
    E_V  &=&  \left< \frac{\rho V}{m} \right>, \\
    E_\textrm{rot} &=& - \left< \psi^* \boldsymbol{\Omega} \cdot \mathbf{J} \psi \right>, 
\end{eqnarray}
where the brackets denote the average over the total volume, and each $E_i$ is actually an energy density per unit volume (i.e., the energy corresponds to the product of these quantities by the total volume). Each corresponds to different fluid-like energy components: $E_k$ is the kinetic energy, $E_q$ is the quantum energy (resulting from the contribution of the quantum pressure), $E_p$ is the internal energy that accounts for particle interactions in the system, $E_V$ is the potential energy associated to the external trap, and $E_\textrm{rot}$ is the contribution from the rotation. Using the Helmholtz decomposition we can further decompose the kinetic energy by separating $(\sqrt \rho \mathbf{v}) = (\sqrt \rho  \mathbf{v})^{(c)} + (\sqrt \rho \mathbf{v})^{(i)}$, where the super-indices $c$ and $i$ denote compressible ($\boldsymbol{\nabla} \times \sqrt \rho \mathbf{v}= 0$) and incompressible ($\boldsymbol{\nabla} \cdot \sqrt \rho \mathbf{v} = 0$) components respectively. With this decomposition we can then define the compressible ($E_k^{(c)}$) and incompressible ($E_k^{(i)}$) kinetic energies in the same way as it is done for classical compressible flows \cite{Nore1997}, such that $E_k = E_k^{(c)} + E_k^{(i)}$.

The total Hamiltonian can be written in a different way using that $-\psi^* \boldsymbol{\Omega} \cdot \mathbf{J} \psi = - \rho \mathbf{v}\cdot (\bm{\Omega} \times \mathbf{r})$. By summing and subtracting $\rho (\bm{\Omega} \times \mathbf{r})^2/2$ inside the integral in Eq.~(\ref{eq:HRGPE}), we obtain
\begin{equation}
    \mathcal{H}[\psi,\psi]= \int d^3 r' \left[ \frac{\rho}{2} |\mathbf{v}_\textrm{rot}|^2 + \frac{\hbar^2}{2 m^2} \left| \boldsymbol{\nabla} \sqrt{\rho} \right|^2 + \frac{g \rho^2}{2m^2} + \left( \frac{V(\mathbf{r})}{m} - \frac{(\bm{\Omega}\times \mathbf{r})^2}{2}\right) \rho\right].
    \label{eq: rot hamiltonian expanded}
\end{equation}
As a result, we can also define a kinetic energy associated to the velocity in the rotating reference frame $\mathbf{v}_\textrm{rot}$, given by $E_k^{(\textrm{rot})} = \langle \rho |\mathbf{v}_\textrm{rot}|^2/2 \rangle$. Using Eq.~(\ref{eq:vrot}) and the fact that $\mathbf{v}$ and $\mathbf{v}'$ are the same fields passively rotated, from the laboratory reference frame the velocity $\mathbf{v'}$ in the equilibrium must be such that it minimizes its distance to the solid body rotation $\bm{\Omega} \times \mathbf{r}'$, which provides a minimum of $|\mathbf{v}_\textrm{rot}|^2$ and of $\mathcal{H}$ in the rotating frame. However, as $\mathbf{v}$ is irrotational everywhere but in topological defects, quantum vortices must be present in the equilibrium. This results, for strong enough rotation, in the appearance of a triangular vortex lattice, the so-called Abrikosov lattice \cite{Abrikosov:1956sx}. Finally, from the last term in Eq.~(\ref{eq: rot hamiltonian expanded}) note that the centrifugal force affects the trapping potential in the direction perpendicular to $\boldsymbol{\Omega}$. In particular, for a harmonic potential of the form $V(\mathbf{r})= m \omega_{\perp}^2 (x^2 + y^2)/2$, and for $\boldsymbol{\Omega} = \Omega \hat{z}$, we can write $(\bm{\Omega} \times \mathbf{r})^2 = \Omega^2 (x^2 + y^2)$. Therefore, rotation in a harmonic trap causes a shift in the effective frequency of the external potential of the form $\omega_{\perp}^2 \rightarrow \omega_{\perp}^2 - \Omega^2$. This means that in a harmonic trap, when $\Omega > \omega_{\perp}$ the condensate will no longer be contained by the potential. These observations will define the three regimes of rotating quantum fluids discussed next.

Note also that using Eq.~(\ref{eq:vrot}), the Hamiltonian in Eq.~\eqref{eq: rot hamiltonian expanded} can be written in a form similar to that of a charged particle in an external magnetic field. This similarity can be exploited to understand some properties of rapidly rotating BECs as was done in \cite{Fletcher2021, Mukherjee2022}.

\subsection{Regimes of rotating BECs \label{sec:regimes}}

As seen in the previous section, rotation has a pronounced effect on the energy minimum of the system. For sufficiently large angular velocities inertial forces become relevant, but the quantization of circulation and the shift in the harmonic potential frequency lead to distinct regimes depending on the rotation rate. At low rotation speeds, when $\Omega$ is smaller than some critical value $\Omega_c$ for vortex excitation, no vortices are present, or those excited externally display no large-scale order. At an intermediate regime with $\Omega_c < \Omega < \omega_\perp$, a well-ordered vortex lattice becomes dominant. Finally, at very large rotation rates such that $\Omega > \omega_\perp$, the system can enter the lowest Landau level (LLL) regime \cite{Fetter_2009}, resulting in the depletion of density in the centre of the condensate and in strong modifications in its dynamics. The system as a whole constitutes a condensed matter problem coupled to the disordered gas dynamics. We therefore consider first these regimes at the equilibrium or with small perturbations from their corresponding equilibria.

\subsubsection{Slow rotation ($\Omega \lesssim \Omega_c$)}

The condensate cannot generate vortices with less than one quantum of circulation. As a result, there is a critical $\Omega$ for the system to start rotating. At this critical value the condensate can create one vortex at its centre, in such a way that its velocity corresponds to the system rotating as a rigid body. Estimation of this threshold is a non-trivial question by itself. In the case with $\boldsymbol{\Omega} = \Omega \hat{z}$ and of an harmonic and axisymmetric trap with $V(\mathbf{r}) = m \omega_{\perp}^2(x^2 + y^2)/2$, using the Thormas-Fermi approximation (i.e., neglecting the kinetic energy of the BEC) a critical value for a single vortex to be both locally and globally stable is given by \cite{Fetter_2009}
\begin{equation}
    \Omega_c = \frac{5\hbar}{2mR_{\perp}^2(\Omega)} \ln \left[ \frac{R_{\perp}(\Omega)}{\xi}\right],
\end{equation}
where $\xi = [\hbar^2/(2g\rho_c)]^{1/2}$ is the condensate healing length (of the same order as the vortex core size), $R_{\perp}(\Omega)$ is the radius of the condensate in the direction perpendicular to $\boldsymbol{\Omega}$ which in the Thomas-Fermi limit is given by $R_{\perp}^2(\Omega) = \hbar \sqrt{2 g \rho_c} /[m \xi (\omega_{\perp}^2-\Omega^2)]$, and $\rho_c$ is the mean density at the centre of the trap. Despite the fact that a thermodynamic critical value for stability can be computed, there is evidence in experiments and simulations that the first appearance of vortices occurs at a rotation $\bar{\Omega}_c>\Omega_c$ \cite{Madison2000, Madison2001, AboShaeer2001}. For slightly larger rotation speeds a few vortices appear in the system, generating a small ordered array that minimizes the energy. Small arrays of vortices have been studied in liquid helium and BECs in \cite{Yarmchuk1979, Coddington2003}. 

If we slightly perturb the system from the equilibrium, restitutive forces allow for the excitation of waves. For small $\Omega$, as none or very few ordered vortices are present, the most relevant waves are the ones seen in non-rotating condensates. In the absence of rotation and without an external potential ($V=0$), GPE can be linearized around an equilibrium with uniform mass density $\rho_0$. The resulting normal modes are sound waves, that satisfy the Bogoliubov dispersion relation
\begin{equation}
    \omega_B(k) = ck \sqrt{1+\frac{(\xi k)^2}{2}}, 
\end{equation}
where $c=(g \rho_0/m^2)^{1/2}$ is the sound speed. Note that the healing length controls the minimum wavelength for the dispersion relation to be approximately linear \cite{Pethick2008}. In the following sections, when dealing with BECs in harmonic potentials or in other inhomogeneous trapping potentials, we will estimate typical values of $c$ and $\xi$ using the central density $\rho_c$.

If a quantum vortex is present in the BEC, small deformations also result in wave excitations. These correspond to Kelvin waves, named after Lord Kelvin, who first derived them for classical fluids in the late 19th century \cite{Thomson1880}. When $\Omega=0$, their dispersion relation is
\begin{equation}
    \omega_K(\mathbf{k}) = \frac{2 c \xi}{\sqrt{2}r^2_n} \left(1 \pm \sqrt{1+k_{\parallel} r_n \frac{K_0(k_{\parallel}r_n)}{K_1(k_{\parallel r_n})}} \right),
\end{equation}
where $r_n$ is the radius of the vortex core, $K_0$ and $K_1$ are modified Bessel functions, and $k_{\parallel}$ is the wave number along the vortex direction. The value of the vortex radius estimated from theoretical arguments or simulations is typically $r_n \approx 2\xi$ \cite{Nore1997, ClarkdiLeoni2015}. This dispersion relation is the same as the classical one derived by Lord Kelvin, but based on the quantum of circulation $\Gamma_0$ instead of the circulation associated to the total vorticity in a classical vortex. When $\Omega \neq 0$ rotation shifts the Kelvin wave dispersion relation. For a single quantum vortex in the rotating frame it becomes
\begin{equation}
    \omega_K^{(\textrm{rot})} (\mathbf{k}) = \Omega + \omega_K (\mathbf{k}) .
\end{equation}

Finally, along with these waves, if a large-wavelength velocity field is excited, we can have inertial waves as in a classical fluid, in which the restitutive force is the Coriolis force in the rotating reference frame \cite{Cambon_1997}. And in the presence of a confining potential, a global oscillation of the BEC can be excited. This is the breathing mode, which in a harmonic potential has frequency $2\omega_\perp$, and that corresponds to a periodic expansion and contraction of the gas in the trap \cite{Pethick2008}.

\subsubsection{Rapid rotation ($\Omega_c < \Omega < \omega_\perp$)}

For rotation sufficiently larger than $\Omega_c$, many quantum vortices may be present in the system. As the condensate tries to adjust to the solid body rotation, a dense uniform vortex array develops if the BEC is big enough---such that inhomogeneity can be neglected. The density of vortices per unit area, $n_v$, must follow from the fact that the total circulation generated by $\mathcal{N}_v$ parallel quantum vortices crossing a closed surface with area $\mathcal{A}$ is $\Gamma = \Gamma_0 \mathcal{N}_v  = \Gamma_0 n_v \mathcal{A}$. To match the rotation, this circulation must equal the solid body circulation in the same area, $\Gamma_{sb} = 2\Omega \mathcal{A}$. Then \cite{Feynman1955}
\begin{equation}
    n_v = \frac{m\Omega}{\pi \hbar} .
\end{equation}
In this regime the fundamental state is two-dimensional---it has translation symmetry along the direction of $\boldsymbol{\Omega}$. We can also estimate the mean radius of a circular cell occupied per vortex as \cite{Fetter2008},
\begin{equation}
    l = \sqrt{\frac{\hbar}{m\Omega}}.
\end{equation}
Consequently, the inter-vortex distance in the vortex lattice, $\ell_\textrm{int}$, in this regime is $\approx 2 l$.

As discussed previously, rotation changes the trapping potential, decreasing the effective frequency of a harmonic potential. That makes the BEC expand---remember that in the Thomas-Fermi approximation, $R_{\perp}^2(\Omega) = \sqrt 2 \hbar c/[m \xi (\omega_{\perp}^2-\Omega^2)]$. As a result, and unlike the homogeneous case, the number of vortices will grow faster than linearly with the rotation. In fact, the total number of vortices in the BEC grows approximately as 
\begin{equation}
n_v \pi R_{\perp}^2 = \left( \frac{R_{\perp}}{l} \right)^2 \propto \frac{\Omega}{(\omega_{\perp}^2-\Omega^2)^2}.     
\end{equation}

As a result of the interaction between vortices, a stable lattice cannot have an arbitrary shape. Vladimir Konstantinovich Tkachenko studied arrays of straight parallel vortices in infinite domains in two papers \cite{Tkachenko1965, Tkachenko1966}. His work has been highly praised in general---despite not publishing more than around $10$ papers--and called ``a \textit{tour de force} of powerful mathematics'' by Freeman Dyson \cite{Sonin2014}. While his approach was very general, he was particularly interested in the case of rotating superfluid helium. He computed the energy of arbitrary vortex lattices, and showed that the triangular lattice has the lowest energy. This is compatible with observations in type II superconductors, in superfluid helium and BECs \cite{Andereck1982, Peretti2023, AboShaeer2001, Coddington2003}, and even recently in vortex arrays in classical rotating fluids \cite{ClarkdiLeoni2015}.

When the equilibrium is perturbed, besides the waves already discussed for non-rotating and slowly rotating condensates, now the lattice can sustain waves that correspond to global modes of oscillation. These modes are called Tkachenko waves, as they were first described by him in 1966 \cite{Tkachenko1966}. For the triangular lattice in a harmonic trap, using elasticity theory, they satisfy
\begin{equation}
    \omega_T^2(k) = \frac{2 C_2}{\rho_c} \frac{c^2k^4}{4 \Omega^2+ [ c^2 + 4(C_1+C_2)/\rho_c ]k^2},
\end{equation}
where $C_1$ is the compressional modulus and $C_2$ is the shear modulus of the vortex lattice \cite{Baym2003}. There are two Thomas-Fermi limits that can be considered for these waves: the rigid limit when $\Omega$ is small compared to the lowest compression frequency $ck_0$ (where $k_0$ is the wave number of the fundamental mode in trap), and the soft limit when $\Omega$ is larger than $c k_0$ but smaller than $mc^2/\hbar$ and compressibility cannot be neglected---i.e., compression of the BEC becomes relevant in response to lattice deformations. In the rigid limit, with low compressibility ($\xi \Omega /c \ll 1$), $C_1 = - C_2$, and a series expansion of $C_2$ yields the dispersion relation \cite{Baym2003}
\begin{equation}
    \omega_T^{(r)} = \sqrt{ \frac{\Omega \xi c}{2^{3/2}} } k,
\end{equation}
while in the soft limit the dispersion relation has the form ($\gamma \approx4$) \cite{Baym2003}
\begin{equation}
    \omega_T^{(s)}=\left[ \left(1 - \gamma \frac{\sqrt{2}\Omega \xi}{c} \right) \frac{\xi c^3}{8\sqrt{2}\Omega} \right]^{1/2}k^2 .
\end{equation}

\subsubsection{Very fast rotation: Low Landau level ($\Omega \gtrsim \omega_\perp$)}

For very fast rotation, the nature of a gaseous BEC differs significantly from that of liquid helium, mainly because of its larger compressibility. In quantum gases, the vortex core size depends strongly on rotation. This is the effect of the centrifugal force, or equivalently, of the frequency shift of the trap, which causes the condensate central density to decrease as mass moves outwards, which in turn modifies the healing length---remember that $r_n \approx 2\xi$. At the same time, the density of vortices in the lattice increases with $\Omega$, packing the vortices closer together. This effect has been studied for harmonic traps when $\Omega < \omega_\perp$ using the Thomas-Fermi approximation \cite{Fetter_2009}.

When $\Omega / \omega_\perp \approx 1$, the vortex core size becomes comparable to the inter-vortex distance, and the Thomas-Fermi approximation cannot be used any more. In this regime, as the result of the strong rotation and of the gas expansion, the inter-particle interaction can become small enough that the whole system becomes two dimensional with a new effective Hamiltonian \cite{Fetter2008}. This Hamiltonian is, for one particle, formally the same as that of a charged particle in an external magnetic field. It is well known that for strong magnetic fields---or, in our case, for a trapped particle under very fast rotation---the energy levels are quantized and exhibit degeneracy, in the so-called Landau levels \cite{Fetter2008, Cohen-Tannoudji2019-au}. For low gas densities and not too strong rotation, most particles will accumulate at the lowest Landau levels (LLLs). The energy gap between two levels is $2\hbar \omega_\perp$, and the mean interaction energy is proportional to $g \rho/m$, which is assumed small in this limit. The relevant parameter for quantifying the system behaviour then is $g \rho/(2\hbar m \omega_\perp)$ which, in experiments, have been explored up to values of $\approx 0.6$ \cite{Fetter2008}. For large rotation speeds, the density in the BEC is large only in a circular ring, with an almost empty centre. If rotation is increased further, a quantum Hall regime can be met, and for which the GPE approximation breaks down \cite{Fetter2008}.

\subsection{Studies of rotating quantum turbulence \label{ref:literature}}

In recent years, numerous studies—both experimental and numerical—have directly or indirectly addressed the topic of rotating quantum turbulence in systems ranging from liquid helium to ultracold quantum gases. These works have produced a wide variety of results, including diverse spectral scaling laws, vortex line decay rates, and more. However, a unifying framework remains elusive. In this section, we take initial steps toward constructing such a framework, one that accounts for the intricate interplay between equilibrium configurations—associated with the various condensed matter regimes—and turbulent dynamics. We focus on rotation as a natural order parameter, as it gives rise to qualitatively distinct states through the influence of inertial forces and the emergence of long-range ordered structures, hallmarks of rotating quantum fluids. Our goal is to lay out a set of conceptual building blocks and propose a perspective that integrates existing results and can serve as a foundation for future developments. To that end, we begin by reviewing relevant studies in the field, distinguishing—due to fundamental physical differences—between results in liquid helium and those in atomic BECs.

\subsubsection{Liquid helium}

Since the pioneering experiments of William F.~Vinen in the mid-20th century, significant progress has been made in the study of turbulence in liquid helium. The Abrikosov vortex lattice has motivated numerous investigations of rotating superfluid helium. However, only very recently turbulence flows were excited in this regime. Early studies of rotating helium primarily focused on vortex nucleation, the formation of equilibrium Abrikosov lattices, and their stability \cite{Yarmchuk1979, ruutu1997}.

Further advances in the generation and measurement of quantum vortices required substantial improvements in laboratory techniques. Abrikosov lattices were successfully created in $^3$He under weakly interacting conditions in \cite{Lounasmaa1999}. More recently, experimental evidence of vortex lattices in $^4$He was presented in \cite{Peretti2023}. In that study, thermal forcing induced collective wave propagation and enhanced vortex interactions, which the authors interpreted as a possible pathway toward turbulence. More recently, in another ${}^4$He experiment using a counterflow configuration, the rapid amplification of helical Kelvin waves was observed, ultimately developing into a turbulent state \cite{Dwivedi2024}. The authors report a family of power-law decay exponents for the vortex line density (see table \ref{table:literature}). In another effort, a recent study in $^3$He-B reported a turbulent regime driven by large-scale inertial waves, which are believed to transfer energy to smaller scales and trigger a cascade of Kelvin waves \cite{Mkinen2023}.

From the theoretical and numerical point of view, a central challenge when studying turbulence in superfluid helium is the extreme separation of the relevant length scales. For instance, the healing length is $\xi \approx 10^{-10}$ m in ${}^4$He and $\xi \approx 10^{-8}$ m in ${}^3$He-B, while the inter-vortex distance is typically $\ell_\textrm{int} \approx 10^{-5}$ m, and the system size is often of the order of centimetres (i.e., $\approx 10^{-2}$ m) \cite{Barenghi2014}. This disparity makes direct numerical simulations of the full system impossible, even if the dynamic equations were known. As a result, a hierarchy of models has been developed, each targeting a different scale of the problem. On the microscopic level, GPE, though originally derived for weakly interacting Bose gases, has been qualitatively successful in capturing some features observed superfluid helium turbulence. However, GPE fails to account for the roton minimum in the excitation spectrum---a limitation that can be addressed through suitable modifications as discussed in \cite{Mller2020}. At mesoscales, vortex filament models have proven particularly useful in describing superfluid dynamics, especially in ${}^4$He \cite{Schwarz1988, Hnninen2014}. In the rotating case, such models were used to study rotating bundles of vortices \cite{Tsubota2003}. Beyond these, a variety of other models have been proposed, which are comprehensively reviewed in \cite{Barenghi2014}.

\subsubsection{BECs and numerical simulations of GPE and RGPE}

In the last two decades, since the first observations of turbulence in atomic BECs in the laboratory in 2009 \cite{Henn2009, Henn2009PRL}, the rapid advance of controlling techniques in experiments and of computing power in simulations allowed for a detailed exploration of multiple regimes, and of quantum turbulence in dilute trapped gases under many different conditions \cite{Barenghi2023}. The focus of many of these studies has been on characterizing the system behaviour, developing methods to stir the condensate out of equilibrium, finding similarities and differences with superfluid helium, and identifying the similarities and differences with classical turbulence. The effects of changing the system dimensionality, as the flows excited change from three-dimensional to two-dimensional, were considered in \cite{Gauthier_2019, Johnstone_2019, Muller_2020}. This exploration---together with motivations ranging from condensed matter to astrophysics \cite{Shukla2024, Liu2025, Sivakumar2025}---led in recent years to the consideration of rotating BECs, or of flows in atomic BECs under some externally created rotation, as the presence of the Abrikosov lattice results in an approximate two dimensionalisation of the system, allows for the excitation of collective modes, and introduces significant constraints in the flow dynamics. This sparked an interest on how out-of-equilibrium rotating BECs differ from classical rotating turbulence, and on how the waves observed in rotating classical fluids are modified by the new normal modes available in the system and by the large-scale order imposed by the lattice.

Some of the first simulations to report turbulent flows in atomic BECs which in some cases have a net rotation were reported by in 2017 by Cidrim et al.~\cite{Cidrim2017}, and in 2021 by Marino et al.~\cite{Marino2021}. The authors considered the decay of muti-charged vortices, which depending on the preparation of the initial state, can result in rotating condensates. In other words, the net angular momentum in the initial condition can lead to a set of corotating vortices when an initially multi-charged vortex breaks up. The authors reported a decay of the mean vortex length---which gives an estimation of the largest scales in the flow---with time of the form $L \sim t^{-1}$, and an energy spectrum $\sim k^{-1}$ which the authors interpret as compatible with Vinen turbulence, a regime of turbulence which can only be observed in quantum flows and has no classical counterpart \cite{Barenghi2023}.  In \cite{Cidrim2017} the authors also claimed that as a result of this behaviour, the turbulent energy cascade cannot develop. Instead, in \cite{Marino2021} the authors proposed that a direct cascade of particles develops, with a particle density spectrum $n(k)\sim k^{-3}$. If an energy cascade exists, the authors could not identify it because of a strong oscillatory behaviour observed in the energy, resulting from strong breathing modes excited in the trap.

Later, a numerical study by Hossain et al.~considered rotating turbulence in unitary Fermi gases \cite{Hossain_2022}. The dynamics of rotating fermionic and bosonic superfluids is relevant for rotating neutron stars, and nuclear many-body systems. The authors focused on the differences in the dissipation mechanisms between fermionic and bosonic superfluids, which manifest as a different damping of Kelvin waves in each case. 

A detailed study of scaling laws in quantum turbulence in rotating BECs by explicitly integrating RGPE was reported in 2022 by Amette Estrada et al.~\cite{AmetteEstrada2022a}. The case of a harmonic trap with a freely decaying initial vortex tangle was considered. For rotation frequencies $\Omega$ smaller than the perpendicular frequency of the trap $\omega_\perp$, a non-Kolmogorovian incompressible kinetic energy spectrum was found with a scaling of $k^{-1}$. It was also found that the system develops a negative temperature state with self-organization of the kinetic energy, or in other words, an inverse kinetic energy cascade reminiscent of the one seen in classical rotating turbulence. Also, an almost flat spectrum was reported for the compressible kinetic energy, which is compatible with a one-dimensional thermalisation process of sound waves propagating in the direction parallel to $\boldsymbol{\Omega}$. Superimposed to a breathing mode with frequency $2\omega_\perp$, it was finally shown that soft Tkachenko waves were excited in the system in the perpendicular direction, which provided an anisotropic dissipation mechanism for the energy at small scales. The $\sim k^{-1}$ incompressible kinetic energy spectrum was explained as resulting from the disordered motion of the lattice. 
The Fourier transform of an incompressible field that is generated by an array of vortices can be written as $\hat{\mathbf{u}} = \sum_j e^{-i\mathbf{k}\cdot\mathbf{r}_j}\hat{\mathbf{u}}_v$, where $\hat{\mathbf{u}}_v$ is the Fourier transform of the velocity generated by a single vortex, and $\mathbf{r}_j$ is the position of the $j$-th vortex. The power spectrum of the total velocity is then given by the integration over angles in Fourier space of
\begin{equation}
    \hat{\mathbf{u}}^*\cdot \hat{\mathbf{u}} = \sum_{j,k} e^{-i\mathbf{k}\cdot (\mathbf{r}_j-\mathbf{r}_k)} |\hat{\mathbf{u}}_v|^2,
    \label{eq:amette}
\end{equation}
which, for a field composed by disordered vortices results in the sum of the spectra of each individual vortex, which scales as $\sim k^{-1}$ \cite{Nore1997, Polanco_2021}. 

An exploration of turbulence in the slowly and rapidly rotating regimes was presented by Amette Estrada et al.~in \cite{AmetteEstrada2022b}, by solving RGPE for BECs in cigar traps. The system displays a phase-like transition as $\Omega$ is increased, going from only a direct cascade of kinetic energy to a split cascade with an inverse cascade of kinetic energy. The free decay of the total vortex length was found to be strongly quenched as $\Omega$ increases. Finally, in \cite{AmetteEstrada2022b} it was also shown that, in the freely decaying case, the system decays towards finite temperature Abrikosov lattices.

Binary BECs where considered in a study by Das et al.~\cite{Das2022}, where they performed two-dimensional simulations and created vortices using a rotating paddle in the laboratory frame. They studied different rotation speeds, and consistently found Kolmogorov-like incompressible kinetic energy spectra scaling as $k^{-5/3}$. Notably, in all their simulations, the stirring frequency was very large relative to the trapping potential, and the system was allowed to evolve freely after the stirring ended.  Another work considering a similar system under rotating elliptical deformations was later done by da Silva et al.~\cite{daSilva2023}. In the transient turbulent state---before the system becomes dominated by vortex arrays---the authors also report robust $k^{-5/3}$ scaling in the incompressible kinetic energy spectrum. The two species exhibit consistent behaviour, and their stirring protocol is experimentally feasible, allowing for different $\Omega/\omega_\perp$ values for each species. More recently Tomio et al.~\cite{Tomio2024} also considered binary BECs but with a Gaussian rotating obstacle, obtaining a $\sim k^{-5/3}$ spectrum for the incompressible kinetic energy.

\begin{landscape}
\begin{table}
\centering
\resizebox{\linewidth}{!}{%
\renewcommand{\arraystretch}{2.6}
\begin{tabular}{|l|c|c|c|c|c|c|c|c|}
\hline
Reference & Sim/Exp & Fluid & Vessel & Forced/Initial condition & Damping & Measured spectra & Vortex decay & $\Omega/\omega_\perp$ \\ \hline

Tsubota et al.~(2003) \cite{Tsubota2003}& Sim (3D vortex filament) & He & -- & Perturbed vortex array & Mutual friction & -- & -- & -- \\ \hline

Cidrim et al.~(2017) \cite{Cidrim2017}& Sim (3D GPE) & BEC & Harmonic trap & Multicharged vortex decay & No & $E_k^{(i)} \propto k^{-1}$ & $L(t)\propto t^{-1}$ & -- \\ \hline

Marino et al.~(2021) \cite{Marino2021} & Sim (3D GPE) & BEC & Harmonic trap & Multicharged vortex decay & No & $E_k^{(i)} \propto k^{-1}$, $n(k) \propto k^{-3}$ & $L(t)\propto t^{-1}$ & -- \\ \hline

Amette Estrada et al.~(2022) \cite{AmetteEstrada2022a} & Sim (3D RGPE) & BEC & Harmonic trap & Free decay & No & $E_k^{(i)} \propto k^{-1}$, w/inverse cascade & -- & $0.29-0.56$ \\ \hline

Amette Estrada et al.~(2022) \cite{AmetteEstrada2022b} & Sim (3D RGPE) & BEC & Harmonic trap & Free decay & No & $E_k^{(i)} \propto k^{-1}$, w/inverse cascade & -- & $\approx 0.84$ \\ \hline

Das et al.~(2022) \cite{Das2022} & Sim (2D GPE) & Binary BECs & Harmonic trap & Rotating paddle & No & $E_k^{(i)} \propto k^{-5/3}$, $E_k^{(c)} \propto k$ & -- & $1-8$ (paddles) \\ \hline

da Silva et al.~(2023) \cite{daSilva2023} & Sim (2D RGPE) & Binary BEC & Harmonic trap (elliptic) & Rotating trap & No & $E_k^{(i)} \propto k^{-5/3}, k^{-3}$ & -- & $0.1-0.8$ \\ \hline

Mäkinen et al.~(2023) \cite{Mkinen2023} & Exp & ${}^3$He-B & Cylinder & Modulated rotation & Mutual friction & Inertial $\to$ Kelvin cascade & -- & -- \\ \hline

Peretti et al.~(2023) \cite{Peretti2023} & Exp & ${}^4$He & Square box & Thermal counterflow + rotation & Mutual friction & -- & -- & -- \\ \hline

Sano \& Tsubota (2024) \cite{Sano2024} & Sim (3D RGPE) & BEC & Harmonic trap (elliptic) & Random forcing + rotation & Linear damping & $E_k^{(i)} \propto k^{-2}$, $E_k^{(c)} \propto k^{0.5}$ & -- & 0.7 \\ \hline

Sivakumar et al.~(2024) \cite{Sivakumar2024a} & Sim (2D RGPE) & BEC & Harmonic trap & Rotating paddles & No & $E_k^{(i)} \propto k^{-5/3}, k^{-1}$ & -- & $0.6-1.2$ \\ \hline

Sivakumar et al.~(2024) \cite{Sivakumar2024b} & Sim (2D RGPE) & BEC & Harmonic trap & Cloud merging + spin-up & No & $E_k^{(i)} \propto k^{-5/3} \to k^{-1}$ & -- & $0$, $0.45$ \\ \hline

Dwivedi et al.~(2024) \cite{Dwivedi2024} & Exp & ${}^4$He & Square channel & Thermal counterflow + rotation & Mutual friction & -- & $L \propto t^{-p}$, $0.7 \leq p \leq 1.5$ & -- \\ \hline

Tomio et al.~(2024) \cite{Tomio2024} & Sim (2D GPE) & Binary BEC & Harmonic trap & Rotating Gaussian obstacle & No & $E_k^{(i)} \propto k^{-5/3}$ & -- & $0.8-1.5$ (obstacle) \\ \hline

\end{tabular}%
}
\caption{Selected papers considering rotating quantum fluids in out of equilibrium regimes. The table provides the references, whether simulations (``Sim'') or laboratory experiments (``Exp'') were done, for the numerical simulations the dimensionality (two or three dimensional, respectively 2D or 3D) and the model solved, the working fluid, how the fluid was constrained, whether the system was forced of freely decaying from some initial conditions, whether damping was used and of what type, the measured spectra if available, the decay of the vortex length in time when reported, and $\Omega/\omega_\perp$ when a harmonic trap was used and when the information is available.}
\label{table:literature}
\end{table}
\end{landscape}

Sano and Tsubota \cite{Sano2024} numerically studied three-dimensional RGPE with random forcing at low wave numbers to excite waves, in a slightly deformed trap inspired on laboratory setups. Starting from an initially anisotropic vortex lattice aligned with the rotation axis, they showed that turbulence disrupts the lattice, partially restoring isotropy in the system. The incompressible kinetic energy remains anisotropic for all times and displays a spectrum with $\sim k^{-2}$ scaling, with distinctive scaling in the directions parallel and perpendicular to rotation. The observed scaling is reminiscent of that observed in classical rotating flows \cite{Cambon_1997, BELLET2006, Sen_2012}. The compressible energy spectrum scales as $\sim k^{1/2}$, which is almost flat and compatible with what was reported in \cite{AmetteEstrada2022a, AmetteEstrada2022b}. Fluctuations remain constant for long times in their simulations, and look relatively small, generating small disturbances of the vortex lattice. 

Rotating quantum turbulence was also studied numerically by Sivakumar et al.~\cite{Sivakumar2024a}, by solving two-dimensional BECs using RGPE under harmonic confinement and with a central perturbed potential that acts as paddles. Turbulence is generated by gradually ramping up the rotation frequency of these paddles in two initial conditions: a vortex-free case, and a case with a vortex lattice. The authors study the kinetic energy spectrum and its flux to characterize the turbulent regimes. For vortex-free initial states, they observe Kolmogorov-like $k^{-5/3}$ scaling and a forward cascade of incompressible kinetic energy, especially for large $\Omega$. In contrast, cases with a vortex lattice display a transient Vinen-like $k^{-1}$ scaling, and develop sustained Kolmogorov scaling only at long times and larger rotation speeds. The study reports, for late times, a forward energy cascade, and an inverse particle density cascade. In several of their simulations the condensate has a void in the centre, resulting from the stirring potential and also because in some simulations large values of $\Omega/\omega_\perp$ are explored, potentially driving the system into the LLL regime. The same authors studied in \cite{Sivakumar2024b} the merging of spinning up clouds using two-dimensional RGPE. They found, independently of whether the system has net rotation or not, a transient incompressible kinetic energy spectrum $\sim k^{-5/3}$ followed by $\sim k^{-1}$. Breathing mode oscillations were also observed with, in both cases, vortices appearing in the condensate. 

As an application of rotating BECs, Sivakumar et al.~\cite{Sivakumar2025} recently used the Gross-Pitaevskii-Poisson equation, which is the usual GPE with an added potential to generate self-gravitation of the condensate, to study the collision of two solitons with net rotation. The study found Kolmogorov-like scaling ($k^{-5/3}$ in the inertial range, with a $k^{-3}$ reported in the ultraviolet), and showed that energy is transferred from the vortex-driven flows to the energy associated to the quantum pressure, which eventually suppresses turbulence. The results highlight differences between self-gravitating and harmonically trapped BECs, and may offer some insight into the structure of dark matter halos and phenomena such as pulsar glitches. 

Table \ref{table: simulations} summarizes many of these results. At this point it is worth emphasizing that, beyond the diversity in the reported scaling laws across experiments and simulations, atomic BECs experiments and simulations also have a wide variety of stirring mechanisms. This diversity is associated to the high degree of control reached in experiments in recent years, enabling fundamentally different ways of setting the condensate in motion. Finally, concerning the scaling laws reported, special care must be taken with $k^{-3}$ scaling laws for large $k$, as this scaling can be an artifact originated in the definition of the kinetic energy spectrum \cite{Nore1997, Krstulovic2010}.

\section{A systematic exploration of parameter space \label{sec:results}}

As it becomes evident from the discussion in Sec.~\ref{sec:review}, many distinct regimes have been identified in rotating quantum turbulence in the literature, exhibiting intriguing differences. However, a complete understanding of the properties and mechanisms resulting in the different scaling laws reported in rotating quantum fluids remains elusive. The existing studies have focused on specific aspects of the problem, and numerical results are often difficult to compare given the many different ways to excite the system, and the different parameters considered.

In this section, we aim to provide a new and ordered exploration of the various regimes that can emerge in quantum turbulence, by keeping the set up fixed, and varying just a few physical parameters in a wide range of values. We focus on the case of atomic BECs. As already mentioned, in this case the inherently compressible nature of quantum gases introduces a non-trivial interplay between condensed matter physics and turbulence. We believe, however, that many of the regimes we identify in this section can be also relevant in liquid helium, with the care needed when extending their validity associated to the different interactions and normal modes present in superfluid helium. We will consider the effect of changing the trapping potential, of changing the gas density, and of changing $\Omega$ to explore the three regimes reviewed in Sec.~2\ref{sec:regimes}. As $\Omega$ increases the system tends to form an Abrikosov lattice, as it becomes effectively two-dimensional under rotation. On the other hand, for sufficiently large $\Omega$ the mass distribution can also become strongly inhomogeneous. In this regime the vortex lattice may be absent, or only visible near the the border of the condensate.

\subsection{Numerical simulations}

\begin{table}
\centering
\begin{tabular}{|c|c|c|c|c|c|c|}
\hline
$\Omega / \omega_\perp$ & Ro & M  & I & $R_\perp/L_0$ & $\left<\rho \right>/\rho_0$ & Trap \\
\hline
0.4  & 0.067 & 0.11 & 25.83 & 1.34 & 0.05 & H\\
0.5  & 0.062 & 0.14 & 23.46 & 1.46 & 0.05 & H\\
0.6  & 0.062 & 0.22 & 19.43 & 1.72 & 0.05 & H\\
0.7  & 0.065 & 0.36 & 14.28 & 1.99 & 0.05 & H\\
0.8  & 0.047 & 0.43 & 10.73 & 2.76 & 0.05 & H\\
0.9  & 0.048 & 0.57 & 6.85  & 2.93 & 0.05 & H\\
0.95 & 0.052 & 0.70 & 2.39  & 3.06 & 0.05 & H\\
0.95 & 0.051 & 0.39 & 10.98 & 2.01 & 0.05 & Q\\
1.05 & 0.050 & 0.61 & 3.60  & 2.51 & 0.05 & Q\\
1.1  & 0.053 & 0.74 & 0.63  & 2.58 & 0.05 & Q\\
0.9  & 0.093 & 0.38 & 0.93  & 2.50 & 0.05 & B\\
0.52 & 0.205 & 0.36 & 53.91 & 2.50 & 0.5  & B \\
\hline
\end{tabular}
\caption{Parameters of the numerical simulations. $\Omega/\omega_\perp$ is the ratio of the rotation speed to the harmonic potential frequency; in box traps we use the effective frequency $\omega_\perp^*$ that would be needed to have a harmonic trap with the same radius as the BEC under the Thomas-Fermi approximation. $\textrm{Ro}$ is the Rossby number, $\textrm{M}$ is the Mach number, $\textrm{I}$ is the interaction parameter using $\omega_\perp$ from the harmonic potential or $\omega_\perp^*$ for box traps, $R_\perp/L_0$ is the condensate radius in units of $L_0$, $\left<\rho\right>/\rho_0$ is the averaged mass density in the whole integration domain in units of $\rho_0$, and ``Trap'' indicates whether a harmonic (H), quartic (Q), or box (B) potential is used.}
\label{table: simulations}
\end{table}

We perform direct numerical simulations of Eq.~\eqref{eq: RGPE} using three different types of axisymmetric potentials: harmonic (H), quartic (Q), and box (B) traps. The domain is periodic in the $z$ direction (parallel to the rotation axis) and no potential is applied in that direction; this results in traps which are conceptually analogous to the elongated (cigar-shaped) traps used in some experiments, but in an idealized limit that allows us to isolate the anisotropic effects of rotation from those that would be caused by a trapping potential in $z$. In the harmonic case the trapping potential is given by $V_H(\mathbf{r}) = m \omega_{\perp}^2 (x^2 + y^2)/2$. For the quartic trap we add to $V_H(\mathbf{r})$ a radial quartic component such that $V_Q(\mathbf{r}) = V_H(\mathbf{r}) + 0.05 \, m \omega_{\perp}^2 (x^2 + y^2)^2$, which introduces a small correction that becomes relevant when the condensate expands too much and its radius becomes large. The third case, the box trap, consists of a rigid cylindrical boundary that confines the condensate. This configuration is inspired by recent laboratory experiments of ultracold atomic gases \cite{Navon2021}, and also mimics the rigid walls used in liquid helium experiments.

To generate initial conditions, we use the method described in \cite{AmetteEstrada2022a} to excite a three-dimensional disordered tangle of vortices, integrating the rotating Landau-Ginzburg equation (RLGE) until reaching a stationary state, which provides us with a solution of RGPE which we then let decay freely with Eq.~\eqref{eq: RGPE}. To solve numerically RLGE and RGPE we use a fourth-order Runge-Kutta method in time, and a fully-dealiased Fourier-based pseudo-spectral method to compute spatial derivatives and non-linear terms, with a resolution of $N^3=512^3$ spatial grid points. The equations are solved using the publicly available parallel code GHOST \cite{Mininni2011, Rosenberg_2020}. The domain has size $[-\pi,\pi]L_0 \times [-\pi,\pi]L_0 \times [-\pi,\pi]L_0$, so that the sides have length $2 \pi L_0$ where $L_0$ is a unit length. To accommodate for the non-periodic nature of the momentum operator and of the trapping potentials in the Fourier bases in the $x$ and $y$ directions, we use the same methods as in \cite{AmetteEstrada2022a} (see also \cite{Fontana_2020} for a general discussion of continuation methods for the Fourier basis).

We performed several simulations as summarized in table \ref{table: simulations}. The table reports the rotation speed in units of $\omega_\perp$ which is fixed for all 
H and Q potentials (for box traps, an effective value $\omega_\perp^*$ is used instead, that corresponds to the frequency that would be needed in a harmonic trap to obtain the same Thomas-Fermi radius as the radius of box trap). The table also lists the Rossby number that measures the strength of the Coriolis force to inertial forces in a classical fluid,
\begin{equation}
    \text{Ro}= \frac{U_\textrm{rms}}{2 \Omega R_{\perp}},
\end{equation}
where $U_\textrm{rms} = \sqrt{2E_k/3}$ is the r.m.s.~gas velocity, the Mach number that measures the ratio of the fluid r.m.s.~velocity to the mean sound speed in the condensate $\langle c \rangle$,
\begin{equation}
    \text{M} = \frac{U_\textrm{rms}}{\langle c \rangle}.
\end{equation}
and the interaction parameter that measures the ratio of the interaction energy to the LLLs gap,
\begin{equation}
    \text{I} = \frac{g \rho_c}{2 \hbar m \omega_\perp}.
\end{equation}
Again, in B traps we use $\omega_\perp^*$ to estimate this parameter. Finally, the table also provides the mean gas density in the whole simulation domain $\left< \rho \right>$, the actual radius of the cloud $R_\perp$ estimated as the radius that contains 95\% of the total mass, and the type of potential used---H, Q, or B.

\begin{figure}
    \centering
    \includegraphics[width=1.0\linewidth]{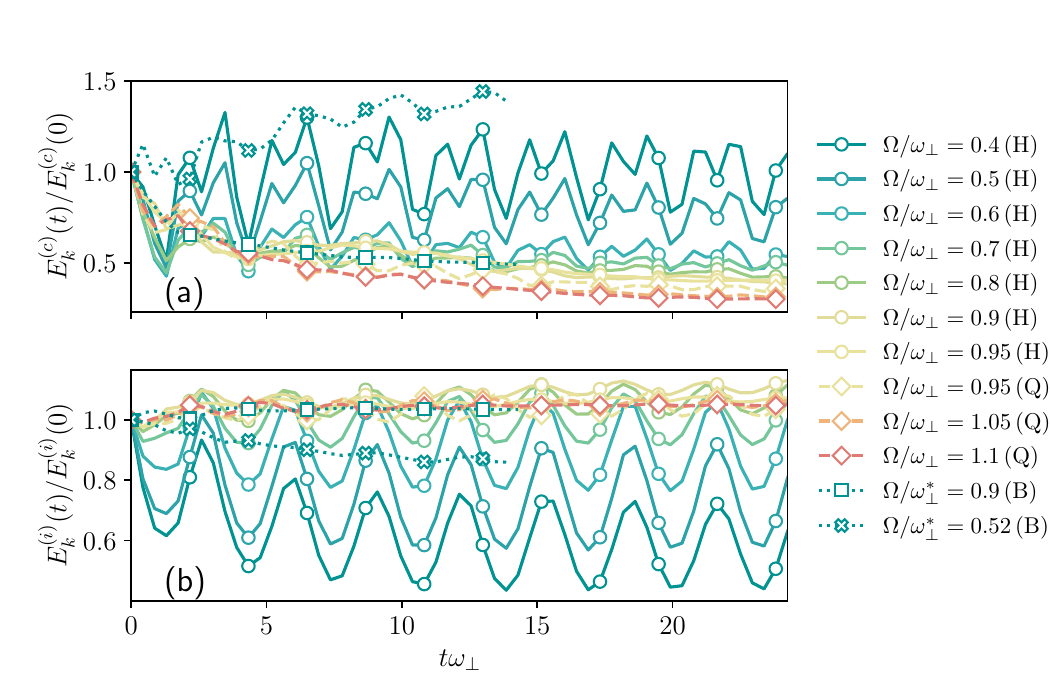}
    \caption{(a) Time evolution of the incompressible kinetic energy in all simulations in table \ref{table: simulations}, normalized by its initial value. Time is multiplied by $\omega_\perp = 1.73 \, U_0/L_0$ in all cases. (b) Same for the compressible kinetic energy.}
    \label{fig:energy_evol}
\end{figure}

Results in the table and in the following sections are presented in units of a characteristic speed $U_0$, the unit length $L_0$ (proportional to the condensate mean radius), and a unit mass $M_0$. All parameters are fixed by setting the reference harmonic trapping frequency to $\omega_\perp = 1.73 \, U_0/L_0$, the speed of sound to $c = 2 \, U_0$, the condensate healing length to $\xi = 0.0088 \, L_0$, and a reference unit mass density to $\rho_0 = 1 \, M_0/L_0^3$. As a reference, and for the sake of completeness, the two simulations with a B trap have the same $\Omega \approx 0.69 \, U_0/L_0$ as the simulation in the H trap with $\Omega / \omega_\perp = 0.4$.  By setting $L_0 = 5 \times 10^{-5}$ m, $U_0 = 2 \times 10^{-3}$ m s$^{-1}$, and $M_0 = 3.2 \times 10^{-20}$ kg, we obtain $\omega_\perp \approx 70$ Hz, $\xi \approx 0.5$ $\mu$m, $c\approx 4 \times 10^{-3}$ m s$^{-1}$, radii of the condensates in table \ref{table: simulations} between $67$ and $150$ $\mu$m, and rotation frequencies between $28$ and $77$ Hz. The mass of the particles is $m \approx 4 \times 10^{-26}$ kg, compatible with $^{23}$Na atoms, and the condensates have $\approx 10^7$ particles. All these values are in good agreement with typical experiments (see, e.g., \cite{AboShaeer2001, White2014}).

\subsection{Global evolution of rotating BECs}

We first discuss the global evolution and main features of all the simulations in table~\ref{table: simulations}, to later consider them based on the different regimes explored and their spectral scaling laws.

\begin{figure}
    \centering
    \includegraphics[width=1.0
    \linewidth]{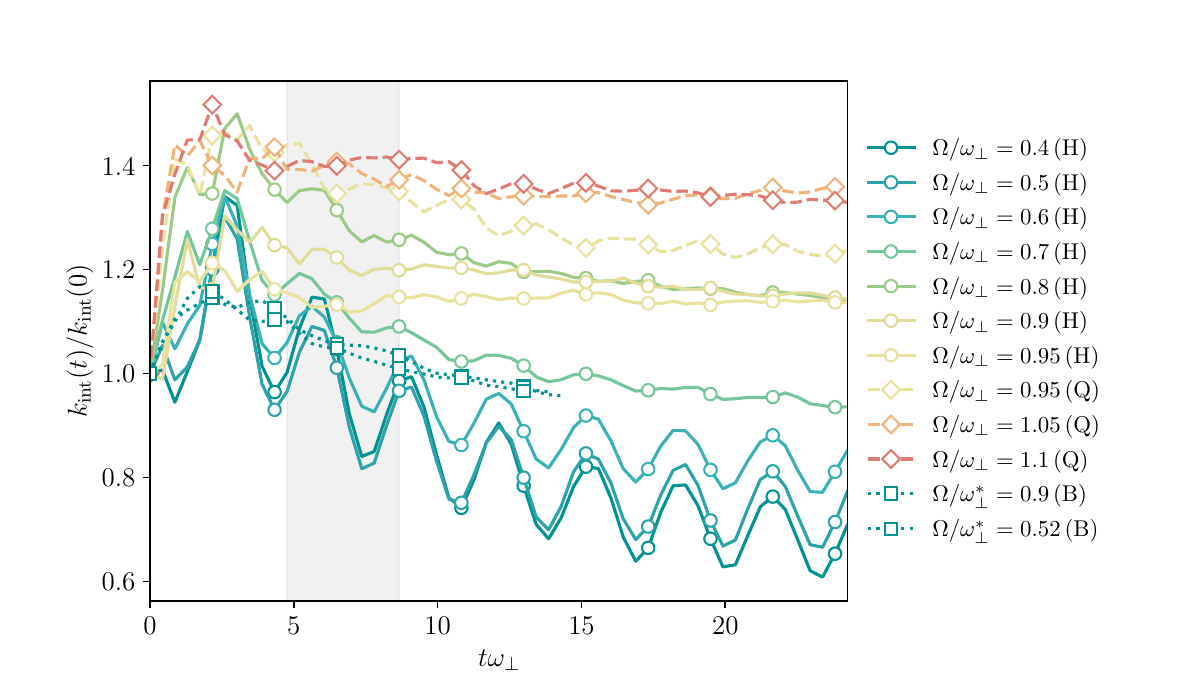}
    \caption{Time evolution of the wave number of the inter-vortex distance, normalized by its initial value, for all simulations listed in table \ref{table: simulations}. Time is given in units of the harmonic trap frequency, $\omega_\perp = 1.73 \, U_0/L_0$. The gray shaded region indicates the time span used for the spectral analysis.}
    \label{fig:kint}
\end{figure}

Figure \ref{fig:energy_evol} shows the time evolution of the incompressible and the compressible kinetic energies in all simulations. Energies are normalised by their initial values, and times are multiplied by $\omega_\perp = 1.73 \, U_0/L_0$ in all cases, such that all quantities become dimensionless. The first conspicuous feature in almost all curves is a global oscillation with frequency $\approx 2 \omega_\perp$. This is the breathing mode of the trap, as the condensate expands and contracts because the gas is not initially at rest. Note that as $\Omega$ increases, the amplitude of this oscillation decreases, and in the case of the box trap these oscillations become negligible or very small.

In Fig.~\ref{fig:energy_evol}(a), in the evolution of $E_k^{(i)}(t)$, note that simulations with small values of $\Omega/\omega_\perp$ display a decay of the incompressible kinetic energy, superimposed to the breathing mode oscillations. However, for $\Omega/\omega_\perp \gtrsim 0.6$ the incompressible kinetic energy stops decreasing, or even slightly increases over time. In the simulations in a box trap (B) both behaviours can be seen depending on the condensate density---which is quantified by the effective value of $\Omega/\omega_\perp^*$ of each BEC in the B trap. As a counterpart, the compressible kinetic energy $E_k^{(c)}(t)$---shown for all simulations in Fig.~\ref{fig:energy_evol}(b)---slightly increases from its initial value when $E_k^{(i)}(t)$ decreases, or decreases in time in all the other cases. This marks a first transition in the behaviour from slow rotation to rapid rotation in the system: as shown in \cite{AmetteEstrada2022b}, when $\Omega \lesssim \Omega_c$ the incompressible kinetic energy is transferred towards smaller scales where it is dissipated as sound waves, while for $\Omega > \Omega_c$ the incompressible kinetic energy stops decaying in time, remains at large scales in the energy spectrum, and this change results in turn in little or no emission of phonons.

\begin{figure}
    \centering
    \includegraphics[width=1.0
    \linewidth]{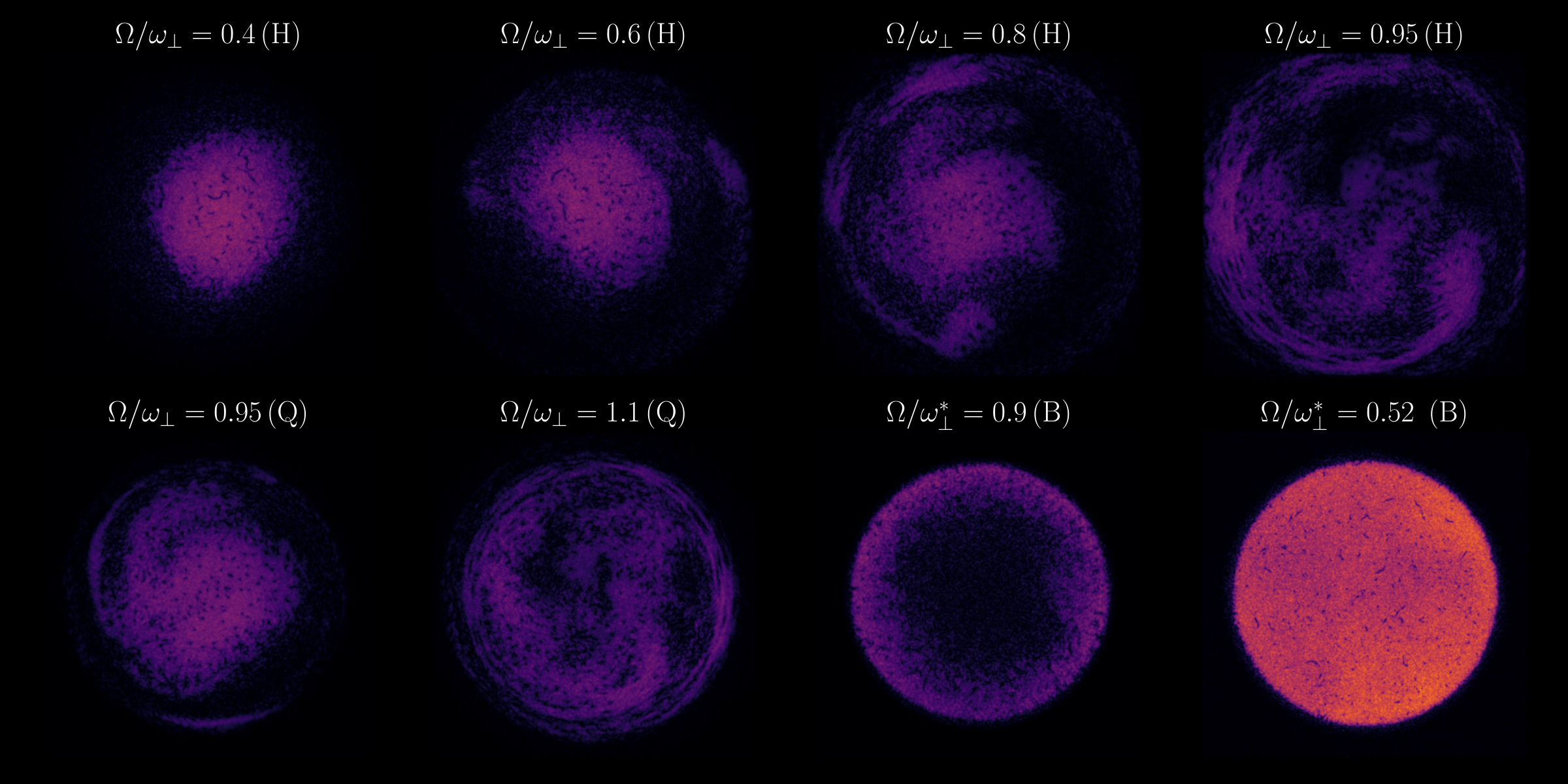}
    \caption{Slices of the mass density $\rho(x,y,z=0)$ for a selection of the simulations in table \ref{table: simulations}. Small black dots and curves correspond to quantum vortices. Cases in H, Q, and B traps are shown. Note how the condensate expands as $\Omega$ increases in the H and Q traps, and how a region of low gas density appears at the centre. In the B trap, note how increasing the gas density results in a more homogeneous condensate with a lower effective $\Omega/\omega_\perp^*$.}
    \label{fig:mass real}
\end{figure}

The behaviour seen in the kinetic energy components has a counterpart in the evolution of the characteristic scales in the flow. The wave-number associated to the inter-vortex distance, $k_\textrm{int} = 2\pi / \ell_\textrm{int}$, as a function of time for all simulations is shown in Fig.~\ref{fig:kint}. This quantity is computed from the spectrum of momentum as described in \cite{Nore1997, AmetteEstrada2022a}. In freely decaying classical turbulence, the maximum of enstrophy---or equivalently, the time of maximum dissipation---is typically used as a reference to identify the time when turbulence is fully developed, and the flow has the largest possible separation between the large-scale eddies and the dissipation scale. Spectra are then often computed at the peak of enstrophy, or by doing a time average for a few eddy turnover times after this peak. For the quantum case, we consider the minimum inter-vortex distance (i.e., the maximum of its wave number) as the time when the smallest scales are excited in the flow.

In Fig.~\ref{fig:kint}, and specially for low rotation frequencies, breathing mode oscillations are again visible in the simulations, specially in the case of the harmonic trap. Note in the figure that large values of $\Omega/\omega_\perp$ result in an early fast growth of the inter-vortex wave number, which then stabilizes and remains approximately constant for a while before slowly decreasing again. This behaviour is compatible with the excitation of multiple collinear vortices in the direction of the axis of rotation \cite{AmetteEstrada2022a}. The resulting lattice also protects the condensate from decaying \cite{Amette_2025}.

As reported previously \cite{Marino2021, AmetteEstrada2022a, Sivakumar2024a}, the strong breathing mode oscillations can significantly affect wave dynamics and the energy spectrum. Averaging over a suitable time window provides a more reliable estimate of scaling laws. Therefore, we will compute spectra shortly after the maximum of $k_\textrm{int}(t)$. As in many cases (specially for harmonic potentials with small rotation) strong breathing mode oscillations are clearly visible in the evolution of $k_\textrm{int}(t)$, in the following we take the first minimum after the peak and average the spectrum over one and a half breathing mode oscillation in the cases with strong variations, and use the same time span for simulations with smaller oscillations. The analysis window used for the spectra is marked in gray in Fig.~\ref{fig:kint}.

\begin{figure}
    \centering
    \includegraphics[width=1.0
    \linewidth]{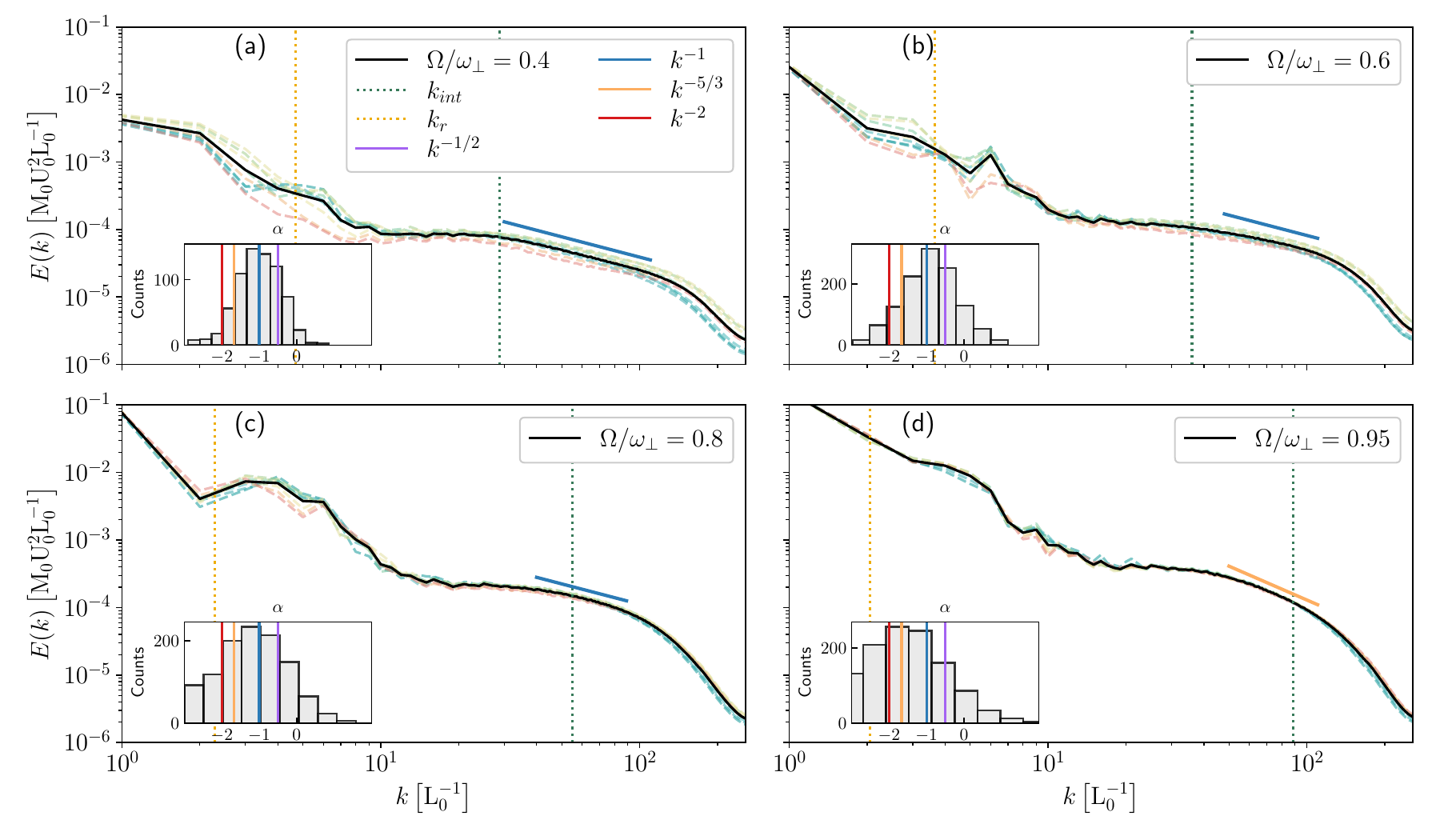}
    \caption{Incompressible kinetic energy spectrum in units of $M_0 U_0^2 L_0^{-1}$, as a function of the wave number in units of $L_0^{-1}$, in four simulations with the H trap (see table \ref{table: simulations}): (a) $\Omega/\omega_\perp = 0.4$. The dashed coloured lines show the spectrum at different times, and the black solid line shows the time averaged spectrum. 
    The inset shows a histogram of the logarithmic derivative of the spectrum (i.e., of the spectral slope). Vertical lines show as a reference spectral indices of $-1/2$ (purple), $-1$ (blue), $-5/3$ (orange), and $-2$ (red). (b) Same for $\Omega/\omega_\perp = 0.6$. (c) $\Omega/\omega_\perp = 0.8$. (d) $\Omega/\omega_\perp = 0.95$. In the main panels, some power laws are shown by solid lines as a reference. Dashed vertical lines indicate the wave number of the BEC radius, $k_r = 2 \pi/R_\perp$, and the inter-vortex distance wave number $k_\textrm{int}$.}
    \label{fig:spectra harmonic}
\end{figure}

To further illustrate the differences between these cases, we examine next a subset of all the simulations. Figure \ref{fig:mass real} shows the mass density $\rho(x,y,z=0)$ for several cases, evaluated at the end of the analysis window marked in Fig.~\ref{fig:kint}, when the system is expected to be in its most disordered state. In the top row, the rotation rate increases from left to right, with all cases using a harmonic trap. The effect of inertial forces is evident: as rotation increases, the condensate expands, and for sufficiently large $\Omega$, the density at the centre of the trap becomes smaller than that in surrounding regions. In the last two panels, a clear ring-like accumulation of mass near the border of the condensate can be seen---signalling that the system is approaching the LLL regime---and in the fastest rotating case, large stretched vortices are also visible close to the edge in this ring.

The first two left panels in the bottom row of Fig.~\ref{fig:mass real} show the density but in the simulations using the quartic trap, with the first having the same rotation speed as the fastest rotating simulation in the harmonic trap, and the second with $\Omega/\omega_\perp >1$. The quartic potential confines better the condensate even at large rotation speeds. The pronounced outer ring of mass concentration seen in the H trap almost disappears in the Q trap with $\Omega/\omega_\perp = 0.95$. However, the case with $\Omega/\omega_{\perp} = 1.1$ shows again a central depletion of mass and a peripheral mass concentration, with the stretched quantum vortices visible again close to the condensate edge.

Finally, the last two right panels in the bottom row of Fig.~\ref{fig:mass real} show the simulations with box potentials. In the first case, which contains the same total mass as all the H and Q trap simulations, the condensate accumulates entirely near the trap border. Note that in the B trap the potential inside the trap is zero, and therefore the centrifugal force can push the condensate outward with no more opposition than the gas pressure. In the last panel, the same simulation but with the total mass increased by a factor of ten is shown. As inter-particle interactions become more relevant, and the gas pressure becomes larger, the condensate fills the trap more homogeneously. This behaviour is also reflected on the smaller value of $\Omega/\omega_\perp^*$ in this case. Overall, these cases highlight how the balance of forces---those resulting from the trap, from rotation, and from particle interactions---are crucial to define what kind of flows the system can sustain.

\subsection{The spectrum of the incompressible kinetic energy}

\begin{figure}
    \centering
    \includegraphics[width=1.0
    \linewidth]{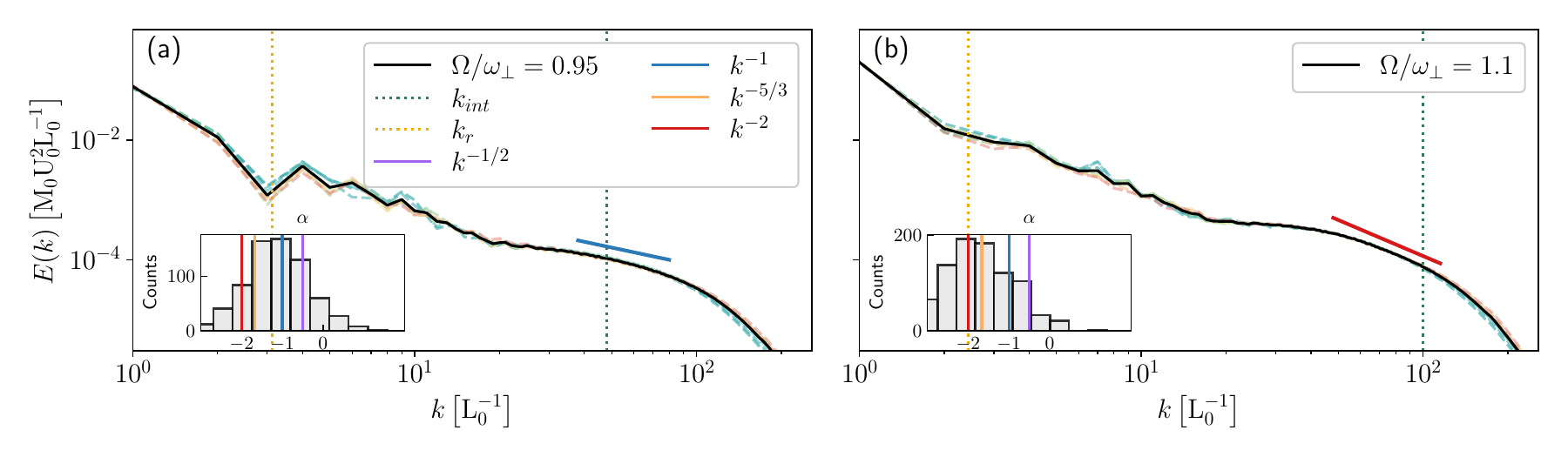}
    \caption{Incompressible kinetic energy spectrum in two simulations with the Q trap  (see table \ref{table: simulations}). The same labels as in Fig.~\ref{fig:spectra harmonic} are used in all panels and insets. (a) $\Omega/\omega_\perp = 0.95$, and (b) $\Omega/\omega_\perp = 1.1$.}
    \label{fig:spectra quartic}
\end{figure}

We now study the properties of rotating quantum turbulence using the spectrum of the incompressible kinetic energy in each simulation. To that end we consider all the spectra at different times, in the time range indicated in Fig.~\ref{fig:kint}. We visually identify ranges of wave numbers compatible with power law scaling, and compute the logarithmic derivative of the spectrum. Note that for a spectrum with an inertial range satisfying the power law $E_k^{(i)}(k) \sim k^{\alpha}$ ($\alpha<0$), then
\begin{equation}
    \alpha = \frac{d \ln E_k^{(i)}(k)}{d \ln k} .
\end{equation}
In practice, as the spectra do not follow a perfect power law, the scaling exponent depends on the wave number and on time, $\alpha(k,t)$. We finally compute histograms of this quantity over a wide range of wave numbers and over time to identify possible scaling laws in each regime of interest.

Figure~\ref{fig:spectra harmonic} shows the incompressible kinetic energy spectrum of four selected simulations with a harmonic potential and with increasing rotation frequency (see table \ref{table: simulations}): The simulations with $\Omega/\omega_\perp = 0.4$, $0.6$, $0.8$, and $0.95$ are each shown in a different panel. 
Each panel displays the spectra at multiple times within the analysis window, along with their time average. Two vertical dashed lines indicate as a reference two wave numbers of interest, the wave number associated to the BEC radius, $k_r = 2\pi / R_\perp$, and the wave number of the inter-vortex distance, $k_{\rm int}$. An inset shows the histogram of the local logarithmic derivative of the spectrum---i.e., of the scaling exponent. Vertical lines in these histograms indicate three reference values, $\alpha = -1$, $-5/3$, and $-2$.

For rotation rates $\Omega/\omega_\perp = 0.4$, $0.6$, and $0.8$ the spectrum is compatible with $\sim k^{-1}$ scaling. This scaling is visible for wave numbers $k>k_\textrm{int}$, i.e., for scales smaller than the inter-vortex distance, with the simulation with $\Omega /\omega_\perp = 0.4$ displaying the smallest value of $k_\textrm{int}$ and thus the broadest inertial range scaling compatible with this power law. However, note that as $\Omega$ increases, a well defined maximum of the histogram of $\alpha$ around $-1$ slowly shifts towards more negative values. Finally, for $\Omega/\omega_\perp = 0.95$, $k_\textrm{int}$ becomes very large with $k_\textrm{int} \xi \approx 0.8$, the peak in the histogram of $\alpha$ is close to $-5/3$, and the incompressible kinetic energy spectrum becomes compatible with Kolmogorov-like scaling for $k \lesssim k_\textrm{int}$. Note that this transition seems to be controlled by how small the inter-vortex distance is, as it approaches the healing length, $\xi$, for increasing $\Omega$.

This transition in the spectrum is associated to the regimes of rotating condensates discussed in Sec.~2\ref{sec:regimes}, and to the underlying dynamics of the system. We have already seen in Fig.~\ref{fig:energy_evol} that in the H trap, as $\Omega$ increases and it becomes such that $\Omega \gtrsim 0.6$, the incompressible kinetic energy stops decaying in time. Now, from the spectrum, we see that as $\Omega/\omega_\perp$ approaches $1$, a new transition takes place in which the spectrum of the incompressible kinetic energy also changes. A key difference between the cases with $\Omega/\omega_\perp = 0.4$, $0.6$, and $0.8$, and the case with $\Omega/\omega_\perp = 0.95$, is that in the latter case the BEC exhibits a strong density depletion in the trap centre, with elongated vortices appearing near the border. Moreover, in this case $\ell_\textrm{int}$ approaches $\xi$, i.e., the inter-vortex distance becomes comparable to the vortex core radius. This suggests that the system is approaching the LLL regime. It is interesting that previous studies with very fast rotation (see, e.g., \cite{Das2022, Sivakumar2024b, Tomio2024} and table \ref{table:literature}) also report similar scaling laws. 

\begin{figure}
    \centering
    \includegraphics[width=1.0
    \linewidth]{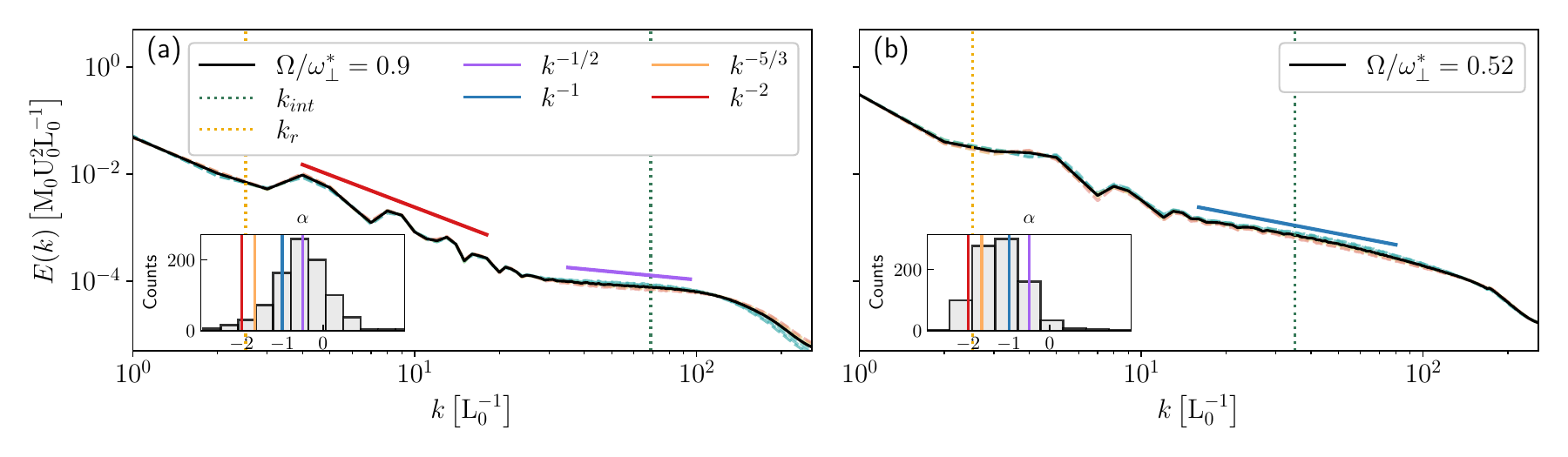}
    \caption{Incompressible kinetic energy spectrum in the two simulations with the B trap. The same labels as in Fig.~\ref{fig:spectra harmonic} are used in all panels and insets. (a) $\Omega/\omega_\perp^* = 0.9$ with $\langle \rho \rangle / \rho_0 = 0.05$, and (b) $\Omega/\omega_\perp^* = 0.52$ with $\langle \rho \rangle / \rho_0 = 0.5$.}
    \label{fig:spectrum_box}
\end{figure}

The differences in the observed spectra can be attributed simply to an effect of this central density depletion---i.e., to the change in the shape of the BEC resulting from the balance of forces in the system. Or it can result from the suppression of vortex lattice dynamics. Previous studies have proposed that vortex lattices and their associated excitations---such as Tkachenko waves---play a crucial role in the dynamics and in the observed scaling laws in the rotating turbulent regime \cite{AmetteEstrada2022a}. However, Tkachenko waves are favoured in configurations with large, coherent vortex lattices, while the lattice is disrupted as the system approaches the LLL regime.

To study these effects, Fig.~\ref{fig:spectra quartic} shows the spectra of two simulations in a Q trap. For $\Omega/\omega_{\perp} = 0.95$ in Fig.~\ref{fig:spectra quartic}(a), which can be compared with the same value of $\Omega/\omega_{\perp}$ in the H trap in Fig.~\ref{fig:spectra harmonic}(d), the spectrum becomes flat again and compatible with $\sim k^{-1}$ scaling---see the histogram of $\alpha$ in the inset. This behaviour is the result of the better confinement provided by the quartic potential, which decreases the radius of the BEC as seen in Fig.~\ref{fig:mass real}, increases the central density, and increases the interaction parameter as seen in table \ref{table: simulations}---i.e., it pushes the system away from the LLL regime. In turn, this allows for the formation of vortex lattices, and for perturbations of the lattice to develop. As rotation increases further in the Q trap, see Fig.~\ref{fig:spectra quartic}(b) for $\Omega/\omega_\perp = 1.1$, the interaction parameter decreases to its smallest values in all simulations with $\textrm{I} = 0.63$, and the energy spectrum steepens again, this time approaching $\sim k^{-2}$ scaling.

Finally, Fig.~\ref{fig:spectrum_box} shows the spectrum of the incompressible kinetic energy in the two simulations with the B trap. The simulation with $\Omega/\omega_\perp^* = 0.9$ and $\textrm{I} = 0.93$ in Fig.~\ref{fig:spectrum_box}(a) displays a possible $\sim k^{-2}$ scaling at small wave numbers, followed by a very shallow spectrum compatible with $\sim k^{-1/2}$ at larger wave numbers---see also the histogram of $\alpha$ in the inset. This simulation has a region of low density in the centre of the trap (see Fig.~\ref{fig:mass real}), as other simulations with large $\Omega$ discussed previously. As the density of the BEC is increased at fixed $\Omega$, resulting in $\Omega/\omega_\perp^* = 0.52$ and $\textrm{I} = 53.91$, $k_\textrm{int}$ becomes smaller and the spectrum shown in Fig.~\ref{fig:spectrum_box}(b) recovers $\sim k^{-1}$ scaling.

\subsection{Scaling laws of rotating quantum turbulence}

\begin{figure}
    \centering
    \includegraphics[width=0.9
    \linewidth]{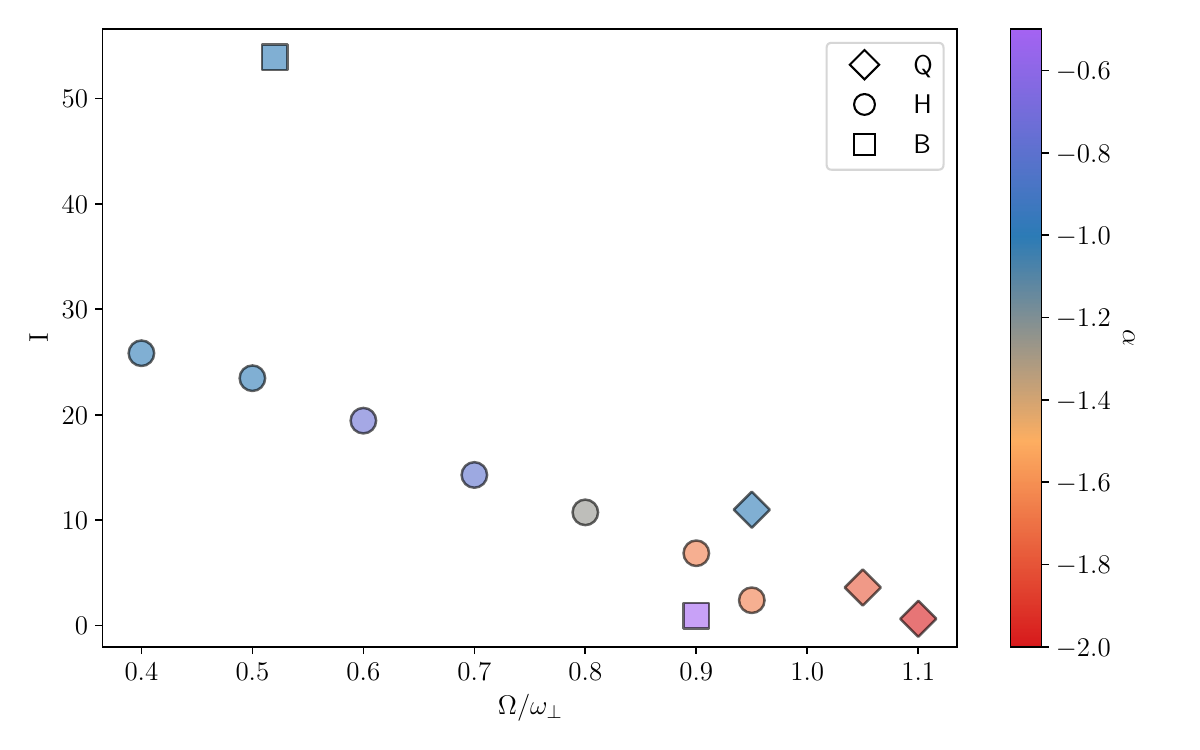}
    \caption{Parameter space of all simulations in table \ref{table: simulations} as a function of $\Omega/\omega_\perp$ and of the interaction parameter $\textrm{I}$. The shape of the symbols indicates the three traps considered (H, Q, and B). The colour of the symbols indicates the value of the spectral index, $\alpha$, obtained from the maximum of the histogram of each simulation.}
    \label{fig:phase_diagram}
\end{figure}

At this point we have explored many different ways in which the spectrum of rotating quantum turbulence can scale. The dominant parameters to separate these regimes seem to be the strength of rotation, measured by $\Omega/\omega_\perp$, and the interaction parameter $\textrm{I}$ that measures the population of the LLLs when small. Together these two parameters can quantify the relevance of inertial forces, the relevance of the vortex lattice in the system dynamics, and how much the condensate is deformed by rotation. Figure \ref{fig:phase_diagram} shows all simulations in table \ref{table: simulations} in the configuration space defined by these two parameters, with the shape and the colour of each marker indicating respectively the type of trapping potential used, and the spectral index of the inertial range.

Most simulations in Fig.~\ref{fig:phase_diagram} lay approximately over a line---at fixed $\omega_\perp$ in the H trap, as $\Omega$ increases the central density decreases, and the interaction parameter $\textrm{I}$ decreases. However, changing the potential to consider the Q and B traps, or changing the mass density in the B trap, allows the exploration of other regions of parameter space. For large interaction parameter $\textrm{I}$, all simulations display $\sim k^{-1}$ scaling. As $\Omega/\omega_\perp$ approaches $1$ and if the potential allows for the central density of the condensate to decrease, $\ell_\textrm{int}$ approaches $\xi$, and $\textrm{I}$ decreases. This results in the population of the LLLs, and a $\sim k^{-5/3}$ scaling is observed instead. Two extreme cases with large $\Omega/\omega_\perp$ and very small values of $\textrm{I}$ display instead a $\sim k^{-2}$ spectrum, followed in the particular case of the B trap by a shallow $\sim k^{-1/2}$ spectrum at large wave numbers.

These results are compatible with several of the previously reported results (see table \ref{table:literature}). However, the parameters $\Omega/\omega_\perp$ and $\textrm{I}$ are not reported in some of the studies, and as a result we can only comment qualitatively on the possible reasons for the agreement or disagreement with the data in Fig.~\ref{fig:phase_diagram}. Most of the studies in the rapid rotation regime \cite{Cidrim2017, Marino2021, AmetteEstrada2022a, AmetteEstrada2022b, Sivakumar2024a, Sivakumar2024b} display $E_k^{(i)}(k) \sim k^{-1}$ scaling as the simulations shown here, even in cases in which the trap geometry, the forcing mechanism,  or the initial conditions differ vastly from the ones considered in our study. In many of these studies, the presence of a vortex lattice was reported, at least at late times after the turbulence decayed---see, e.g., \cite{AmetteEstrada2022b}. Many of the studies in the very fast rotation regime, with $\Omega$ approaching or above the trap frequency, display $\sim k^{-5/3}$ scaling \cite{Das2022, Sivakumar2024a, Tomio2024}, also as the simulations in the same regime in this study. In these cases the systems are characterized by the absence of a recognizable vortex lattice, and in many cases by a depletion of mass density in the trap centre. Finally, the study in \cite{Sano2024} reports a spectrum compatible with $\sim k^{-2}$. While this scaling may seem compatible with some of our simulations, a close examination of their results indicates that their regime differs significantly from the one reported here that have strong fluctuations in both the velocity and the density at very fast rotation, while the simulations in \cite{Sano2024} were conducted with $\Omega/\omega_\perp = 0.7$ and the authors report a well-defined vortex lattice. These apparent inconsistencies suggests that this regime warrants further study.

This provides a possible picture of what is happening. In the presence of rotation with $\Omega_c < \Omega < \omega_\perp$, when vortex lattices can develop and $\ell_\textrm{int} > \xi$, the incompressible kinetic energy spectrum is $\sim k^{-1}$. This spectrum can result from disordered motions of the vortex lattice, as discussed in \cite{AmetteEstrada2022a} and in Eq.~(\ref{eq:amette}), or from the non-linear interaction of collective motions of the Abrikosov lattice. The excitation of Tkachenko modes has indeed been confirmed in this regime by studying the spatio-temporal spectrum of the condensate in \cite{AmetteEstrada2022a}. From the theory of wave turbulence \cite{Nazarenko2011}, the expected spectral scaling for a system with $N$-wave interactions and a dispersion relation of the form $\omega \sim k^{\gamma}$ in $d$ dimensions, is given by
\begin{equation}
   \alpha = d - 6 + 2 \gamma + \frac{5 - d - 3\gamma}{N - 1} .
   \label{eq:WT}
\end{equation}
Note that for $d=2$, as excitations of the Abrikosov lattice are mostly two-dimensional, $\gamma = 2$ as $\textrm{Ro}$ is moderately small and thus we expect the lattice to be ``soft''---i.e., to allow for compression modes---and $N=4$ for interactions between four waves in RGPE, we obtain $\alpha = -1$.

As rotation increases with $\Omega \gtrsim \omega_\perp$, $\ell_\textrm{int}$ approaches $\xi$ and the population of the LLLs increases, a transition into a $\sim k^{-5/3}$ spectrum is seen. This Kolmogorov-like scaling has been reported in quantum turbulence either resulting from the reconnection of quantum vortices at wave numbers smaller than $k_\textrm{int}$, or from a Kelvin wave cascade at wave numbers larger than $k_\textrm{int}$. As in the rotating case $k_\textrm{int}$ is large and the scaling is observed for wave numbers $k \lesssim k_\textrm{int}$, and as in this regime closely packed and elongated vortices are seen near the border of the condensate, it may be the case that the observed scaling is the result of vortex reconnection.

Finally, for even larger values of $\Omega$ and small values of $\textrm{I}$ in Q and B traps we see $\sim k^{-2}$ scaling, accompanied in one case by a very shallow spectrum at large wave numbers. The $\sim k^{-2}$ scaling can be the result of the interaction of inertial waves as in classical rotating turbulence \cite{Cambon_1997}, specially when it develops at large scales, or the result of the non-linear interaction of Tkachenko modes. Indeed, from weak turbulence theory and using again Eq.~(\ref{eq:WT}), now with $d=2$ and $\gamma = 1$---i.e., in the ``hard'' limit as $\textrm{Ro}$ is smaller in these simulations, indicating the lattice should be stiffer and less prone to compression---an exponent $\alpha=-2$ is obtained independently of the value of $N$.

\section{Conclusions}

The study of turbulence in rotating quantum fluids reveals a rich and intricate phenomenology that departs significantly from both classical rotating turbulence as well as from homogeneous and isotropic quantum turbulence. Through a review of the literature and a systematic numerical exploration of the relevant parameter space---namely, rotation rate, interaction strength, and system size---we have identified three distinct regimes: a slowly rotating regime in which the Abrikosov lattice does not develop and turbulence retains features reminiscent of the isotropic case, a rapidly rotating regime characterized by strong vortex lattice order and anisotropic dynamics, and a low Landau level regime where the system starts to enter into a quantum-Hall-like phase which can be dominated by collective modes and quantized excitations, but which can also display a classic-like regime at large scales in some circumstances. The different turbulent scaling laws of the energy spectrum depend both on how fast the system rotates, and how strong interactions are in the BEC. We can also expect some differences in these regimes depending on whether gaseous BECs or liquid helium are used in experiments, as some regimes may not be accessible in the liquid helium case.

These regimes are compatible with the diverse scaling laws and flow dynamics previously reported in the literature, even in cases in which the trap geometry or the stirring mechanism were different. Their existence signals a shift in how turbulence should be understood in quantum systems subject to rotation. Unlike the case of classical fluids, where rotation modifies the energy transfer in an otherwise continuous turbulent cascade, rotation in quantum fluids can drive the system into qualitatively different states of quantum matter. As a result, explaining the dynamics in these regimes requires more than a traditional turbulence framework. Instead, it calls for a synthesis of tools from turbulence theory and many-body quantum physics. The interplay between quantum vortices, collective excitations, and rotation-induced order introduces mechanisms that are absent in classical systems. As a result, rotating quantum turbulence stands as a paradigmatic example of strongly interacting, out-of-equilibrium physics where fluid dynamics and condensed matter theory must converge to provide a description of the system. 

This work aimed at providing a roadmap for understanding this convergence, pointing toward future investigations that may uncover universal principles governing turbulence in quantum systems with external constraints. Several questions remain open for the future. We would like to point out three in particular. First, a detailed comparison between rotating flows in BECs and in superfluid helium would be interesting to pinpoint the different regimes accesible in each case. Second, a detailed analysis of the transfer mechanisms and of the role of the different waves in each regime could shed more light into the turbulent cascades. Finally, changing the potential or the total mass in the BEC would allow for a more complete exploration of the configuration space, to identify other regimes, or the transition between the regimes reported here.

\ack{J.A.E.~and P.D.M.~acknowledge support from UBACyT Grant No.~20020220300122BA, and from proyecto REMATE of Redes Federales de Alto Impacto, Argentina.}

%%%%%%%%%% Insert bibliography here %%%%%%%%%%%%%%

\bibliographystyle{RS}
\bibliography{sample}

%\begin{thebibliography}
%\end{thebibliography}

\end{document}